# GaAs nano-ridge laser diodes fully fabricated in a 300 mm CMOS pilot line

Yannick De Koninck[1,a], Charles Caer[1,*], Didit Yudistira[1,*], Marina Baryshnikova[1], Huseyin Sar[1], Ping-Yi Hsieh[1,2], Saroj Kanta Patra[1,b], Nadezda Kuznetsova[1,c], Davide Colucci[1,3], Alexey Milenin[1], Andualem Ali Yimam[3], Geert Morthier[3], Dries Van Thourhout[3], Peter Verheyen[1], Marianna Pantouvaki[1,d], Bernardette Kunert[1,*], Joris Van Campenhout[1,*]

[1]imec, Kapeldreef 75, Leuven, 3001, Belgium.
[2]KU Leuven, Kapeldreef 75, Leuven, 3001, Belgium.
[3]Photonics Research Group, Ghent University-imec, Technologiepark-Zwijnaarde 126, Gent, 9052, Belgium.

[a]Present address: NVidia Corporation, Roskilde, 4000, Denmark.
[b]Present address: AMS-OSRAM International GmbH, Regensburg, 93055, Germany.
[c]Present address: Micro- Nano Systems, KU Leuven, Kapeldreef 75, Leuven, 3001, Belgium.
[d]Present address: Microsoft Corporation, Cambridge, CB1 2FB, United Kingdom.

[*]Corresponding authors: Charles Caer (charles.caer@imec.be), Didit Yudistira (didit.yudistira@imec.be), Bernardette Kunert (bernardette.kunert@imec.be), Joris Van Campenhout (joris.vancampenhout@imec.be)

**Silicon photonics is a rapidly developing technology that promises to revolutionize the way we communicate, compute, and sense the world [1,2,3,4,5,6]. However, the lack of highly scalable, native CMOS-integrated light sources is one of the main factors hampering its widespread adoption. Despite significant progress in hybrid and heterogeneous integration of III-V light sources on silicon [7,8,9,10,11,12], monolithic integration by direct epitaxial growth of III-V materials remains the pinnacle in realizing cost-effective on-chip light sources. Here, we report the first electrically driven GaAs-based multi-quantum-well laser diodes fully fabricated on 300 mm Si wafers in a CMOS pilot manufacturing line. GaAs nano-ridge waveguides with embedded p-i-n diodes, InGaAs quantum wells and InGaP passivation layers are grown with high quality at wafer scale, leveraging selective-area epitaxy with aspect-ratio trapping. After III-V facet patterning and standard CMOS contact metallization, room-temperature continuous-wave lasing is demonstrated at wavelengths around 1020 nm in more than three hundred devices across a wafer, with threshold currents as low as 5 mA, output powers beyond 1 mW, laser linewidths down to 46 MHz, and laser operation up to 55 °C. These results illustrate the potential of the III-V/Si nano-ridge engineering concept for the monolithic integration of laser diodes in a Si photonics platform, enabling future cost-sensitive high-volume applications in optical sensing, interconnects and beyond.**

Silicon photonics enables the miniaturization and mass manufacturing of optical systems for a growing set of applications [13,14,15,16]. However, the lack of native, low-cost coherent light sources is a major roadblock for ubiquitous adoption of the technology, especially for future high-volume cost-sensitive applications, such as chip-to-chip optical interconnects in machine-learning systems [17], fibre-to-the-X applications [18], or optical sensors for consumer devices [19,20]. In many of today's datacom products, the light sources are manufactured and tested separately on their native III-V substrates, and subsequently hybrid integrated on the silicon photonics wafers in the form of micro-assembled laser packages [21] or through high-precision flip-chip assembly [7,8,9]. Owing to the sequential nature and high-precision requirement of such assembly processes, the manufacturing throughput of these integration solutions may not scale to meet the aggressive density, cost and volume targets for future products. To address these challenges, a variety of hybrid and heterogeneous III-V integration techniques with better scaling potential are currently being developed and commercialized. Micro-transfer printing enables parallel, back-end-of-line integration of pre-fabricated III-V components with higher throughput and integration density [22,23,24]. A front-end-of-line alternative is heterogeneous III-V integration, involving the die-to-wafer bonding of non-patterned III-V device layer stacks directly on the silicon photonics wafer with subsequent III-V device patterning and CMOS-based back-end-of-line metal interconnect formation. This approach has been thoroughly developed over the past several years [10,11,12] and is now available in at least two commercial manufacturing lines [25,26]. However, deep cost reductions and wide adoption of this technology in mainstream CMOS fabs may be hampered by the remaining need for die-to-wafer bonding and expensive III-V donor substrates used for epitaxial III-V growth, much of which are disposed as waste during the manufacturing process, raising additional concerns around health, safety and environmental sustainability.

For these reasons, the direct epitaxial growth of high-quality III-V optical gain materials selectively on the desired device locations on large-size silicon photonics wafers remains a highly sought-after objective. Unfortunately, the large mismatch in crystal lattice parameters and thermal expansion coefficients between III-V and Si materials inevitably initiates the formation of crystal misfit defects, which are known to deteriorate laser performance and reliability. To reduce the defectivity in the III-V layers, multiple research groups have developed thick buffer layers and strained superlattice layers with great success [27–29]. In particular, GaAs-based laser stacks featuring InAs quantum-dot (QD) optical gain regions have unlocked tremendous progress in the performance and reliability of monolithically integrated lasers on Si, leveraging the improved tolerance of QDs against residual crystal defects, as compared to conventional quantum-well laser designs [30]. However, these results have been limited to die-level demonstrations only. Thick buffer layers are difficult to implement on large-diameter wafers, as they are prone to formation of cracks and other defects arising from thermally induced stress. Recently, encouraging extensions of this work have been reported, featuring epitaxial growth in deep pockets etched in a silicon-oxide masking layer on 300 mm [31] and 200 mm silicon photonics wafers [32]. However, the molecular-beam epitaxy (MBE) employed for the deposition of the QD gain stack is intrinsically non-selective, resulting in the deposition of a polycrystalline III-V film outside of the targeted trenches, potentially hampering subsequent wafer-scale integration processes.

Selective-area growth (SAG) of III-V materials on patterned Si wafers by metalorganic vapor-phase epitaxy (MOVPE) represents a compelling integration approach, arguably with better scalability potential than MBE. Selectivity can be easily achieved in vapor-phase epitaxy, and as such, the III-V material can be deposited only where it is needed. The confinement of misfit defects is achieved by starting the growth of the III-V layers in deep and narrow trenches, etched in a dielectric masking layer on the Si wafer, a technique referred to as "aspect-ratio trapping" (ART) [33–35]. Various ART architectures are currently being explored and efficient defect reduction has been reported without the need for thick buffer layers [36–38]. Consequently, issues with wafer warpage and crack formation in the wafer-scale deposited III-V layers are much less pronounced. Over the past several years, we have developed the concept of nano-ridge (NR) engineering, a SAG-based integration approach applying ART for defect reduction, followed by the growth of low-defectivity III-V nano-ridges outside of the trenches. By careful tuning of the MOVPE process parameters, the nano-ridge dimensions, shapes and composition can be accurately engineered, independent from the starting trench dimensions. Optimized nano-ridge structures typically feature threading dislocation densities well below $10^5$ cm$^{-2}$, up to 100x lower than for optimized blanket buffer layers (see Methods and Supplementary Information Epitaxy and crystal defects). Previously, we successfully applied the nano-ridge engineering concept to realize a variety of heterostructures, in-situ doping profiles and surface passivation layers in various III-V material systems such as GaAs, InAs, GaSb and InGaAs [39–45], leading to first device demonstrations including optically pumped lasers [46], heterojunction bipolar transistors [47] (HBTs) and wafer-scale GaAs-based p-i-n photodetectors with record-low dark currents on Si [48].

In this paper, we exploit the III-V nano-ridge engineering concept to demonstrate, to the best of our knowledge, the first full wafer-scale growth and fabrication of electrically pumped GaAs-based lasers on standard 300 mm (001) Si wafers, entirely in a CMOS pilot manufacturing line. Leveraging the low defectivity in GaAs nano-ridge structures, the lasers employ an optical gain region based on InGaAs multiple quantum wells (MQW), embedded in an in-situ doped p-i-n diode and effectively passivated with an InGaP capping layer, all monolithically grown on 300 mm Si wafers. Using on-wafer GaAs nano-ridge p-i-n monitor photodetectors for wafer-scale measurements, more than 300 nano-ridge devices were found to lase around 1020 nm wavelength in continuous-wave operation at room temperature, with threshold currents as low as 5 mA, slope efficiencies as high as 0.5 W/A, and up to 1.75 mW total emitted optical power. Furthermore, an early reliability test revealed continuous lasing during at least 500 h of room-temperature operation, with only a mild increase in threshold current and constant slope efficiency. We believe that these results represent a significant milestone towards the manufacturing of scalable and energy efficient light sources for low-cost optical interconnects, optical sensors and beyond.

## Results

### Laser structure and fabrication

Leveraging wafer-scale processes, several thousands of GaAs nano-ridge devices including lasers, photodetectors and test structures are fabricated on a standard 300 mm silicon wafer, as shown in Fig. 1a-c. Each GaAs nano-ridge structure is formed by selective-area epitaxy, initiated with the growth of n-type doped GaAs in a high aspect-ratio trench patterned in the n-type doped silicon substrate (see fig. 1c). The epitaxy process conditions are tuned to form box-shaped structures outside of the trench, with embedded p-i-n heterojunction. The optical gain region consists of three compressively strained InGaAs quantum wells (QW) with 20% In content in the non-intentionally-doped (nid) GaAs layer (fig. 1d), and the nano-ridge is capped by an InGaP passivation layer. The n-type contacts of the GaAs p-i-n diodes are formed by standard Cu metallization and W plugs landing on the heavily n-type doped top surface of the silicon wafer, which is in turn electrically connected to the n-type GaAs epitaxial layer. The p-type contact is formed by a second row of W plugs, punching through the InGaP passivation layer and landing in the p-type doped GaAs top contact layer (see Extended Data Fig. 1 and Methods for fabrication details). To minimize loading effects and improve uniformity during the epitaxy process, the nano-ridge structures are grown in regular arrays with a pitch of 1 micrometer (fig. 1c).

A cross-sectional high-angle annular dark-field (HAADF-) scanning transmission electron microscopy (STEM) image with energy dispersive X-ray spectroscopy (EDS) of a fabricated device is shown in fig 1e, and highlights the In-containing layers, the full passivation of the nano-ridges outside the trenches by the InGaP layer, as well as the embedded InGaAs quantum wells. The HAADF-STEM image in fig. 1f shows the full extent of the nano-ridge cross-section, revealing a well-controlled nano-ridge geometry from the Si-substrate to the top Cu electrode, as well as the top W contact plug piercing through the InGaP layer and landing in the $p^+$-GaAs contact layer. A close-up view of the active region (dark-field (DF-)STEM in fig. 1g shows defect-free InGaAs QWs and GaAs barriers, while a close-up view of the high aspect-ratio trench (DF-STEM, fig. 1h) shows efficient trapping of threading dislocations (TDs) at the bottom of the n-GaAs region near the silicon interface. Finally, the longitudinal HAADF-STEM image in fig. 1i further illustrates the effective trapping of threading dislocations, which remain confined to the bottom of the ART trench.

As part of the wafer-scale fabrication process, laser cavities are formed in the GaAs nano-ridges by dry etching two facets with an angle of 12°, yielding approximately 5 % facet reflectivity (fig. 2a). On one side of the laser cavities, on-chip monitor photodetectors are implemented in the same epitaxial layers as the nano-ridge lasers, while the other facet adjoins a larger etched area enabling out-of-plane optical emission.

A key aspect to achieve lasing in the GaAs nano-ridges resides in optimizing the pitch of the W plugs contacting the p-GaAs layer. For dense contact pitch designs, the strong spatial overlap between the optical mode propagating in the GaAs nano-ridge waveguide and the W plugs induces very large optical losses that prohibit lasing. Yet, efficient lasing operation is obtained for laser designs with a relatively large W plug pitch ($p_{CON35}$=4.8 µm). In these designs, mode beating of the fundamental $TE_{00}$ and the higher-order $TE_{02}$ mode results in a periodic interference pattern that minimizes the optical intensity locally underneath the W plugs, as illustrated by the simulated mode profiles shown in fig. 2c-d. This mode-beating behaviour enables low-loss optical propagation through the NR laser structure (fig. 2c), while in the case of a tighter W pitch (fig. 2e) ($p_{CON35}$=0.3 µm), the optical field decays quickly due to the strong absorption from the W contacts (see Methods and Supplementary Information for further discussion regarding p-contact plug pitch optimization, calculated mirror reflectivity and simulated internal losses). In addition, the strong optical confinement in the NR, inferred from the 2D finite difference (2D FD) calculations, yields a confinement factor $\Gamma$ between the 3 QWs and the optical mode as large as 8 %, which is instrumental for compensating the large intracavity total optical loss at threshold:

$$\Gamma g_{\text{th}} = \alpha_i + \alpha_m \quad (1)$$

where $g_{th}$ is threshold gain, $\alpha_i$ is the laser internal and $\alpha_m$ is the total mirror loss.

The test configuration used for the wafer-scale characterization of the devices is shown in fig. 2b: three electrical probes are deployed to simultaneously drive the nano-ridge lasers and probe the photodetectors, while a multi-mode fibre (MMF) optical probe is used to collect part of the upwards radiated laser emission, for subsequent power and spectral analysis. Alternatively, as depicted in fig. 2f, one laser facet can be cleaved to form a mirror with approximately 40 % reflectivity, while the other facet remains facing the nano-ridge photodiode. This configuration enables more efficient collection of the laser emission, using a butt-coupled single-mode lensed fibre, allowing for in-depth analysis of the laser performance at the die level.

**Die-level laser results**

First, we discuss die-level laser measurements to unambiguously demonstrate laser operation in the cleaved-facet GaAs nano-ridge devices. In a first measurement, we drive the nano-ridge laser with a continuous-wave current at room temperature and collect the laser output from the cleaved facet by a large-area photodetector. Fig. 3a shows the measured optical output power and diode voltage as a function of laser drive current, i.e. the L-I-V plot for a cleaved device with a cavity length of 1.16 mm. The L-I curve reveals a threshold current as low as 4.5 mA, and a single-facet output power of up to 0.7 mW at a bias current of 20 mA. The dashed line represents the calculated L-I response, obtained by solving the laser rate equations (see Methods and Supplementary Information). The model accurately predicts the measured threshold current and is in reasonable agreement with the measured slope efficiency. The V-I curve shows a diode characteristic with turn-on voltage of 1.4 V. Above the turn-on voltage, a non-linear V-I response can be observed, originating from the non-ohmic and relatively high contact resistance in the widely spaced contact plugs to the p-type GaAs layer (see Supplementary Information).

To gain additional insights into the nano-ridge laser properties, we measured a similar, 1.4mm long laser in a different test configuration (fig 2f), where the large-area photodetector is replaced by a lensed single-mode fibre (SMF) connected to an optical spectrum analyser (OSA). The optical spectra recorded as a function of laser current are displayed in fig. 3b, showing single-mode laser emission around 1023 nm across a wavelength span of 50 nm. A zoom-in of the same spectra (see fig. 3c) shows a threshold current $I_{th}$ of 6 mA and a redshift of the laser wavelength up to 0.4 nm without any mode hopping with increasing drive current. To confirm that the device is indeed operating as a laser, we also carried out a linewidth measurement using a self-homodyne setup (see Methods and Supplementary Information for further reading). From the measured radiofrequency (RF) beat note depicted in fig. 3d, we infer a linewidth of 46 MHz at a current bias $I_{bias}$ of 15 mA ($I_{bias}>2I_{th}$). Finally, the inset in fig. 3d illustrates a side mode suppression ratio (SMSR) exceeding 30 dB at this drive current, ensuring that the RF beat note originates from a single longitudinal mode of the nano-ridge laser.

**Wafer-level laser results**

Next, we discuss the fully wafer-scale fabricated lasers with two etched facets. As a first example, a nano-ridge laser with a 2 mm long cavity was probed on the wafer and the optical emission was measured using the on-chip photodetector (PD) on one side and a MMF on the other side. The I-I plot depicted in fig. 4a shows a threshold current of 7.5 mA and a maximum photocurrent of 71.5 µA at a bias of 17.5 mA. By considering the measured PD responsivity of 0.65 A/W [48] and the simulated LD-to-PD coupling efficiency of 12.5% (see methods), we infer a total emitted power of up to 1.76 mW, and a 1.33 % wall-plug efficiency, which agrees well with the laser model. Subsequently, we carried out full 300 mm wafer-scale measurements to assess the reproducibility of our nano-ridge laser results, using the on-wafer photodetectors. As depicted in figure 5, we measured three sets of nano-ridge lasers with different lengths (fig. 5a: L=1 mm, fig. 5b: L=1.5 mm and fig. 5c: L=2 mm) across a full wafer, gathering statistics regarding process variability and length dependence of the laser threshold, output power and slope efficiency. The performance metrics of more than 300 functional nano-ridge lasers were extracted and are displayed in fig. 5a-c. The calibration previously performed on single nano-ridge devices is used to estimate the total optical power radiated by the two facets of the lasers. Mean threshold current values of 5.9 mA, 8.1 mA and 9.3 mA were measured for 1 mm, 1.5 mm and 2 mm long lasers, respectively. The mean slope efficiencies are 8.3 µA/mA, 6.1 µA/mA and 4.9 µA/mA, or 0.33 W/A, 0.22 W/A and 0.19 W/A, respectively. These measured values are in good agreement with predicted values using the laser model. For L=1 mm, the minimum threshold current is smaller than 6 mA for a large fraction of lasers and as low as 4.5 mA (1.2 kA/cm$^2$), while the estimated maximum total output power exceeds 1.25 mW at a bias current lower than 10 mA, reaching a wall-plug efficiency of 3.3%. For L=1.5 mm, the minimum measured threshold current is 5.5 mA while the maximum total output power exceeds 1.5 mW at a bias current lower than 15 mA. Finally, for L=2 mm, the minimum measured threshold current is 7.1 mA (0.93 kA/cm$^2$) while the maximum total output power exceeds 1.75 mW at a bias current of 18 mA.

The relatively large spread in measured threshold current, slope efficiency and output power across the wafer originates from variability of the wafer-scale processes, which have not yet been optimized for uniformity: first, variations in GaAs NR width and height across the 300 mm wafer originating from the MOVPE process variability induce changes in the confinement factor $\Gamma$ within the InGaAs quantum wells, as well as in the mode-beating periodicity, which in turn increases the absorption loss from the metal contact. Second, wafer-scale variation of the InGaP/GaAs etch process needed to land the W vias on the p-GaAs layer and residual topography after the planarization steps required for the back-end-of-line (BEOL) processing may lead to shorted or open-circuit electrodes. The wafer-scale variability is further illustrated in fig. 5f, which depicts the distribution of the mean threshold current per die for 1-mm long lasers (see also Extended Data Fig. 3). The lowest threshold current values are found mostly in a ring-like shape at half the diameter of the 300 mm wafer. We note that this wafer map is made after binning of the tested devices, resulting in the removal of the dies at the edge of the wafer that contained only light-emitting diodes or short circuits, due to the failure modes described above.

The temperature dependence of the nano-ridge lasers was investigated by mounting the tested wafer on a stage with a temperature controller and recording the L-I curves (fig. 4b). For a 2 mm long NR laser, single-mode lasing was observed at a temperature up to 55 °C, with a threshold current of 30 mA at 55 °C, compared to 7.5 mA at 25 °C. A fit of the temperature dependent threshold current and slope efficiency yields characteristic temperatures $T_1$=29 K and $T_2$=94.9 K, respectively (See Extended Data Fig. 2).

As an early investigation into the reliability of the nano-ridge lasers, we carried out an on-wafer stress test on one of the best performing, 2-mm long nano-ridge lasers. We biased the laser at 1.5 $I_{th}$ at room temperature and monitored the output power, slope efficiency and threshold current (fig. 4c, see Methods for additional information regarding the stress conditions). While the threshold current increased from 6.1 to 7.3 mA, the degradation showed a decelerating trend and the slope efficiency stayed largely constant. The NR laser remained operational during at least 500 h of stress time, exceeding the previously reported record lifetime of 200 hours for GaAs-based QW lasers directly grown on silicon [49].

**Discussion**

To the best of our knowledge, the presented work marks the first monolithic III-V laser diode integration on 300 mm Si wafers entirely carried out in a CMOS manufacturing pilot line, without the need for any III-V substrates or bonding steps. High-quality active GaAs waveguides are grown by MOVPE applying nano-ridge engineering, mitigating any major issues related to wafer bow or crack formation as typically encountered in blanket III-V layers epitaxially grown on silicon. Despite their sub-micrometre-scale cross-sectional dimensions, the GaAs/Si nano-ridge structures feature efficient carrier injection and generate strong optical gain at 1-micrometre wavelengths, enabled by the record-low crystal defectivity in epitaxially grown GaAs on Si and effective passivation by the InGaP capping layer. These attributes are instrumental for robust steady-state laser operation at room temperature and above, as demonstrated in this paper with threshold current densities comparable with conventional laser designs and below 1 kA/cm$^2$. In the demonstrated nano-ridge lasers, an optical multi-mode beating effect mitigates part of the optical losses induced by the top metal contacts, which is equally essential to achieving the low threshold currents and slope efficiencies as high as 0.5 A/W. Furthermore, the grating structure formed by the periodic metal plugs enables single-mode laser operation, with linewidths comparable or smaller than those realized in typical vertical-cavity surface-emitting lasers (VCSELs).

A key remaining challenge is to demonstrate the long-term reliability of the monolithic GaAs-on-silicon nano-ridge lasers. It is well known that quantum-well based lasers on silicon are particularly sensitive to the presence of threading dislocations, and a reliable quantum-well laser essentially needs to have a dislocation-free active region [50]. As shown in the supplementary materials, we estimate the threading-dislocation density in our GaAs nano-ridges to be below 6x10$^4$ cm$^{-2}$, or equivalently, on average one dislocation per 4 mm of nano-ridge length. As such, many of the tested lasers are expected to be dislocation free, providing a positive outlook for achieving long reliability lifetimes. Several devices have indeed shown encouraging room-temperature operating lifetimes beyond 100 h, and the dominant failure mode observed so far is related to the high current density (> 150kA/cm$^2$) in the top metal plugs, which need to be sparsely populated to realize low optical loss in the current laser designs. An in-depth reliability analysis is ongoing and will be reported elsewhere.

We see multiple opportunities to go beyond the current demonstration. First, the nano-ridge cross section can be improved to decouple the optical modes from the metal contacts, enabling a much higher density of p-type contact plugs while retaining low internal optical loss. This will lower the operating voltage and improve wall-plug efficiencies, while also reducing current density and electrical resistance in the top metal contacts, leading to improved reliability. Next, the laser operating wavelength can be extended towards the O-band by growing nano-ridges with higher In-content [51] or alternatively using InAs quantum dots as the optical gain material. Last, the nano-ridge lasers can be optically coupled with (amorphous) silicon waveguides [52]. Coupling the GaAs nano-ridge to silicon and silicon nitride waveguides enables the implementation of external-cavity diode lasers, with better controlled laser wavelengths and reduced laser noise. Such GaAs-Si-SiN integrated photonics platform is expected to enable future cost-sensitive applications in optical interconnects, optical sensing and beyond.

**Fig. 1 | Wafer-scale integration of GaAs nano-ridge lasers on Si. a**, Photograph of a fabricated 300 mm Si wafer containing thousands of GaAs devices. **b**, Close-up view of a fabricated 300 mm wafer showing multiple dies. **c**, Cross-sectional scanning electron micrograph (XSEM) of a GaAs nano-ridge array after epitaxy, before encapsulation in oxide. **d**, Sketch of the cross-section of a GaAs nano-ridge device highlighting the various layers, including the InGaAs quantum wells, p-i-n diode, InGaP passivation layer and metal contacts. **e**, Energy dispersive X-ray spectroscopy (EDS) image of a nano-ridge cross-section highlighting the In-containing layers. **f**, High-angle annular dark-field (HAADF-) scanning transmission electron microscopy (STEM) picture of a transverse cut of a GaAs nano-ridge device. **g**, Dark-field (DF-)STEM close-up view of the InGaAs quantum wells embedded in the unintentionally doped (uid)-GaAs. **h**, DF-STEM close-up view of the n-GaAs/n-Si interface confirming aspect ratio trapping (ART) of threading dislocation (TD) defects. **i**, HAADF-STEM picture of a longitudinal cut of a nano-ridge device, showing an isolated W plug contacting the p-GaAs layer.

**Fig. 2 | GaAs Nano-ridge laser test cell with on-wafer photodetectors for wafer-scale characterization. a**, Bird's eye view of a nano-ridge structure, showing the Fabry-Perot laser formed by two etched facets, the inline photodetector on the bottom right, and the common ground electrode on the bottom left. Dummy GaAs NRs and the encapsulating oxide are omitted in the drawing for clarity. Inset: Cross-section HAADF-STEM image of an etched GaAs facet. **b**, 3D drawing of the wafer-scale test configuration depicting light radiated upwards by the left facet and collected by an MMF and 3 electrical probes driving the NR laser and monitoring the NR PD. **c**, 3D FDTD simulated electric-field intensity plot at wavelength of 1030 nm, projected in the yz plane, for a device with a W top-contact pitch of 4.8 µm. Multi-mode beating is observed with a period of 1.6 µm, minimizing the optical-field intensity at the W contacts and enabling low optical propagation losses. **d**, $TE_{00}$ and $TE_{02}$ electric-field mode profiles at and in-between the W plug locations. **e**, 3D FDTD simulation of a NR device with a dense W contact pitch of 0.3 µm, showing a rapid field decay induced by strong optical absorption loss. **f**, 3D drawing similar to b, showing the measurement configuration for a cleaved-facet device, with a single mode fibre collecting the laser output and 3 electrical probes.

**Fig.3 | Die-level measurement of a single-cleaved-facet GaAs nano-ridge laser. a**, Room temperature L-I-V curves of a 1.16 mm long nano-ridge laser, showing good agreement between measurements (blue solid line) and model (black dashed line), and revealing a threshold current of 4.1 mA and a single-sided slope efficiency of 0.047 W/A. The I-V curve (orange solid line) shows a diode turn-on voltage of 1.4 V. Inset: microscope image of the die-level test setup. **b**, Optical output spectra of the tested laser for bias currents ranging from 3 to 20 mA, showing single-mode emission across a wavelength span of 50 nm. **c**, Close-up view of the same spectra centred around 1023.6 nm, highlighting the absence of mode hopping. Inset: intensity map showing single-mode mode-hop-free lasing with a 0.4 nm redshift of the laser wavelength. **d**, Measured RF power spectrum (blue line) of the tested laser at a bias current of 15 mA, with a superimposed Voigt fit yielding a linewidth of 46 MHz. Inset: Optical output spectrum of the laser showing a side-mode suppression ratio (SMSR) exceeding 30 dB.

**Fig.4 | On-wafer measurement of an etched-facets GaAs nano-ridge laser. a**, L-I-V curves of a 2-mm long nano-ridge laser, showing solid agreement between the measurement (blue solid line) and the model (black dashed line), and indicating a threshold current of 7.5 mA and a slope efficiency of 6.45 µA/mA (0.25 W/A). The I-V curve (orange solid line) reveals a diode turn-on voltage of 1.5 V. Inset: microscope image of the wafer-level test configuration, showing a MMF and 4 electric probes. **b**, Measured fibre-coupled L-I curves of a 2-mm long GaAs laser for different temperatures, showing CW lasing up to 55 °C (see also Extended Data Fig. 2). **c**, Early reliability test, showing continuous room-temperature laser operation up to 500 h at a bias current of 9.17 (1.5$I_{th}$) mA, with a 20 % increase in threshold current, 40 % relative power change at fixed bias and no significant change in the slope efficiency.

**Fig.5 | Full wafer-scale measurements of etched-facets GaAs nano-ridge lasers.** I-I and L-I plots from 300 functional nano-ridge lasers using the on-wafer photodetectors for measuring laser output, for **a**, 1-mm long lasers, **b**, 1.5-mm long lasers, and **c**, 2-mm long lasers. The right y-axis indicates the total output power radiated by the two laser facets. **d**, Measured threshold-current distribution across the wafer. **e**, Measured slope-efficiency distributions for the three laser lengths. In **d** and **e**, each grey dot represents a value extracted from a single device. The red dots represent the threshold current and slope efficiency for different laser lengths as predicted by the laser rate-equation model. **f**, Wafer map showing the mean threshold current per die for 1 mm long nano-ridge lasers (see also Extended Data Fig. 3).

## Methods

### Laser modelling

The mode profile, field overlap with the quantum wells and total internal loss were simulated considering the dimensions of fabricated nano-ridge devices, using the 2D finite-difference (FD) method with Ansys Lumerical software. Optimization of the W plug pitch to minimize propagation loss, and estimation of etched mirror reflectivity and fibre collection efficiency was carried out using 3D finite-difference time-domain simulations in Ansys Lumerical software. The electrical injection efficiency and carrier loss rates were simulated using the technology computer-assisted design (TCAD) Poisson solver in Synopsis Sentaurus software. Finally, the laser threshold current and slope efficiency were computed using Python-based codes solving the standard laser rate equations, enabling us to correlate the simulated parameters with the measured data. We express the general form of these equations as [53]:

$$\frac{dN}{dt} = \frac{\eta_i I}{q V_a} - AN - BN^2 - CN^3 - v_g g_0 \log\left(\frac{N}{N_{tr}}\right) N_p$$

$$\frac{dN_p}{dt} = \left(\Gamma v_g g_0 \log\left(\frac{N}{N_{tr}}\right) - \gamma_p\right) + \Gamma \beta B N^2 ,$$

where $N$ and $N_p$ are the carrier and photon densities, $\eta_i$ is the carrier injection efficiency extracted from the TCAD simulation, $I$ is the injected current, $q$ is the electron charge and $V_a$ is the active volume. $A$, $B$ and $C$ are the Schockley-Read-Hall recombination, the bimolecular recombination and the Auger coefficients, respectively. $v_g$ is the group velocity, $g_0$ the gain coefficient and $N_{tr}$ the transparency carrier density. $\Gamma$ is the confinement factor, i.e., the fraction of the optical mode confined in the quantum wells, $\gamma_p$ is the cavity photon loss rate and $\beta$ is the spontaneous emission factor. The different parameters and their respective values are detailed in Supplementary Information.

### Device fabrication

The devices are processed on 300 mm silicon wafers with {001} orientation, starting with the formation of 300 nm tall Si ridges and subsequent oxide deposition and planarization following a standard shallow trench isolation (STI) process typical in CMOS foundries, and n-type ion implantation of Si. The Si ridges are subsequently wet etched using tetramethylammonium hydroxide (TMAH) to form V-shaped trenches in the STI oxide, with two {111} facets used as the starting surface for subsequent III-V epitaxy. GaAs p-i-n nano-ridges containing three $In_{0.2}Ga_{0.8}As$ quantum wells and capped by an $In_{0.5}Ga_{0.5}P$ passivation layer are grown in one process step using metalorganic vapor-phase epitaxy (MOVPE). The Si wafers are then coated with oxide and planarized. Laser mirrors and photodiode facets are subsequently patterned using DUV lithography and III-V dry etch. Next, the wafers are again coated with oxide and planarized. Subsequently, the fabrication steps move to Back-End of Line (BEOL) processing to form W contact plugs to the p-GaAs and n-Si contact layers. Final metallization is done using a standard CMOS Cu Damascene process. Details of the fabrication process and GaAs epitaxy are reported in the Supplementary Information and a high-level process flow is depicted in Extended Data Fig. 1.

### Die-level measurements

The characterized devices were taken from a fully processed 300 mm wafer cleaved into smaller dies. A high-precision cleaving tool was used to form a mirror in the middle of the NR structure. The lasers were then characterized on a die-level probing setup using a large-area PD to extract the single-facet output power, while the light emitted from the other etched facet is collected by the in-line PD. Spectral analysis of the laser emission was carried out by collecting the emitted light using a butt-coupled lensed fibre connected to an optical spectrum analyser.

**Laser linewidth measurements**
The laser linewidth was measured on a cleaved sample using a self-homodyne interferometric setup (see Supplementary Information for additional details of the setup), containing a 1-km long fibre delay line in one arm of the fiber-based Mach-Zehnder Interferometer. The resulting signal is recorded by a balanced photodiode and sent to an electrical spectrum analyser (ESA). The power spectral density (PSD) of the laser emission was measured under DC bias current, and the laser linewidth was extracted from a Voigt fit of the measured PSD.

**Wafer-level and Reliability measurements**
Wafer-scale laser characterisation was carried out using a 300-mm semi-automated wafer probe station outfitted with a temperature controller and a MMF to collect the laser emission, enabling the measurement of thousands of devices on a single wafer at high throughput. Early reliability measurements were carried out using the same setup, tracking the performance of one of the best performing GaAs NR laser diodes during a 500-h long stress test. In this test, the laser bias current was kept at 9.17 mA (=1.5 $I_{th}$) and the wafer stage was kept at 25 °C. Bias current sweeps were carried out every 20 h, and L-I plots were recorded to extract threshold current, output power and slope efficiency.

**Estimation of total laser output power**
The die-level measurement configuration allows an estimation of the total optical power radiated by both facets. Let us note the output powers at the cleaved facet and the etched facet to be $P_1$ and $P_2$, and their respective transmission coefficients $T_1$ and $T_2$. $T_1$ and $T_2$ are calculated through 3D FDTD simulations (see Supplementary Information) and are evaluated to be $T_1$=0.45 and $T_2$=0.30, respectively. 3D FDTD simulation is also uses to estimate the coupling efficiency of the laser diode into the monitor photodetector (PD), yielding $T_{laser \to PD}$=12.5 %. This value accounts for the reflection at the PD facet, the beam diffraction at the laser facet and the beam divergence in the etched gap between the two devices. Knowing the photodiode responsivity [47] $R_{PD}$=0.65 A/W, the measured optical power on the cleaved-facet side is related to the measured photocurrent on the etched-facet side as:

$$I_{PD} = \frac{T_2}{T_1} R_{PD} T_{laser \to PD} P_1$$

We can extend this relationship to the wafer-level measurement case, where both facets have equal transmission and reflection (i.e., $T_1 = T_2 = T$ and $P_1 = P_2 = P$), which reduces the relationship to:

$$I_{PD} = R_{PD} T_{laser \to PD} P$$

Finally, we can express the total output power of the NR laser from the photocurrent readout, as depicted in fig. 5:

$$P_{tot} = 2 \frac{I_{PD}}{R_{PD} T_{laser \to PD}}$$

**Data availability**
The measured and presented datasets in this study are available from B.K. or J.V.C. upon reasonable request.

**Code availability**
The evaluation scripts and employed models are available from B.K. or J.V.C. upon reasonable request.

**Acknowledgment**
The authors would like to acknowledge support from the staff of imec's 300 mm pilot line.

**Funding**
This work was funded by imec's industry-affiliation R&D program on Optical I/O.

**Competing interests**
The authors declare no competing interests.

**Authors' contributions**
B.K. and J.V.C. initiated the project. Y.D.K., D.Y., M.P, B.K. and J.V.C. designed the NR devices. Y.D.K. and S.K.P. performed TCAD modelling of the devices. Optical simulations were carried out by Y.D.K., D.Y., and D.C. Definition of the process flow was done by D.Y., M.P., B.K. and J.V.C. M.B., D.C. and B.K. executed NR growth and conducted crystal characterization. Coordination of the wafer fabrication in the 300 mm CMOS pilot line was done by D.Y., N.K., P.V. and M.P. H.S. carried out electrical and optical measurements on die level and wafer level, P.-Y.H. performed early reliability tests. N.K. developed the die level setup and performed initial die level measurement. A.M. developed the III-V etch. A.A.Y. built the self-homodyne interferometric setup, performed linewidth measurements of NR lasers and analysed the results under supervision from G.M. and D.V.T. C.C. conducted theoretical investigation of the NR laser. Y.D.K., C.C. D.Y., D.V.T, B.K. and J.V.C. supervised the project and analysed the results. C.C. and J.V.C. wrote the manuscript with inputs from all the authors.

**Correspondence and requests for materials** should be addressed to Charles Caer, Didit Yudistira, Bernardette Kunert or Joris Van Campenhout.

**Extended Data Fig.1 | High-level process flow of the GaAs nano-ridge device fabrication. a** Starting with a 300 mm silicon wafer (001). **b** Silicon ridge patterning. **c** STI oxide deposition and n$^{++}$-implantation. **d** Si trench patterning. **e** MOVPE growth of GaAs nano-ridges. **f** Oxide deposition and planarization. **g** III-V dry etch, oxide deposition and planarization. **h** Oxide re deposition and planarization. **i** W plug contact formation. **j** Cu damascene metallization.

**Extended Data Fig.2 | Temperature and current-dependent measurements of a 2-mm long single-mode NR laser.** **a** L-I characteristics for temperatures ranging from 25 to 60 °C, showing the large spontaneous emission below threshold. **b** Threshold current as a function of temperature extracted from **a**, yielding a characteristic temperature $T_1$ of 29.0 K. **c** Slope $\eta_d$ as a function of temperature extracted from **a**, yielding a characteristic temperature $T_2$ of 94.9 K. **d** Optical spectra at a current of 35 mA for temperatures ranging from 25 to 60 °C, showing SMSR exceeding 20 dB up to 55 °C, while the laser is at threshold when tested at 60 °C. **e** Laser wavelength as a function of temperature, showing a linear slope of 71.1 pm/K extracted from the optical spectra in **d**. **f** Optical spectra at a temperature of 25 °C for currents ranging from 20 to 35 mA, showing a SMSR exceeding 30 dB. **g** Laser wavelength as a function of current, showing a linear slope of 13.6 pm/mA extracted from the optical spectra in **f**.

**Extended Data Fig. 3 | Wafer maps of the mean threshold current and mean slope efficiency for the different laser lengths.** Wafer maps showing the distribution of the mean threshold current per die for (**a**) 1-mm long (**b**) 1.5-mm long, and (**c**) 2-mm long nano-ridge lasers. Wafer maps showing the distribution of the mean slope efficiency per for (**d**) 1-mm long (**e**) 1.5-mm long, and (**f**) 2-mm long nano-ridge lasers

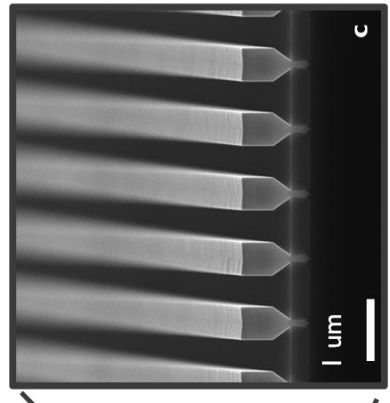
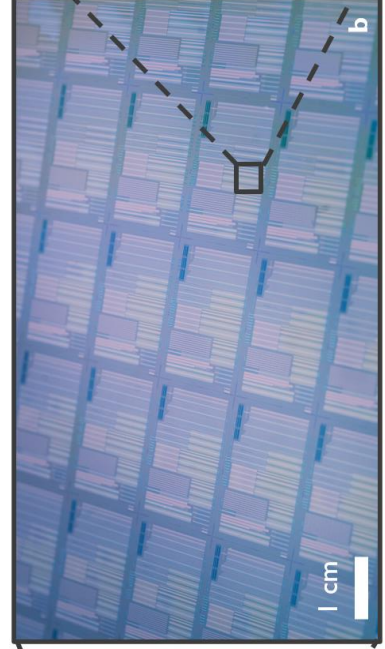
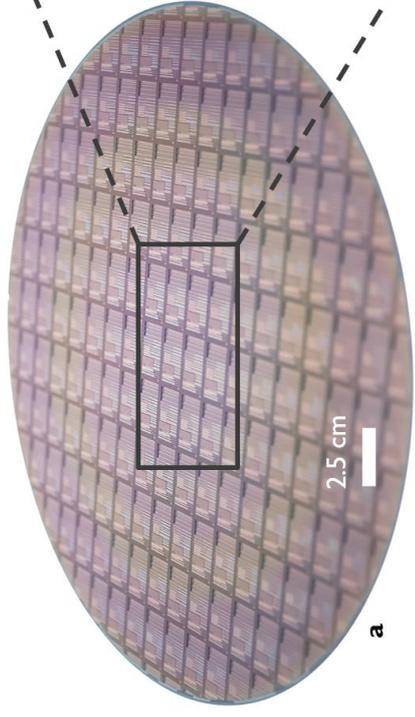
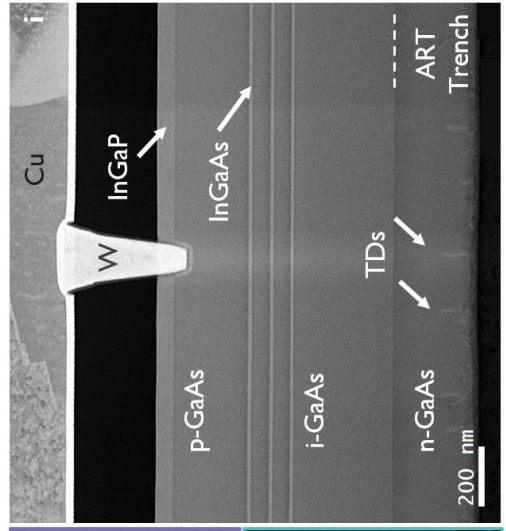
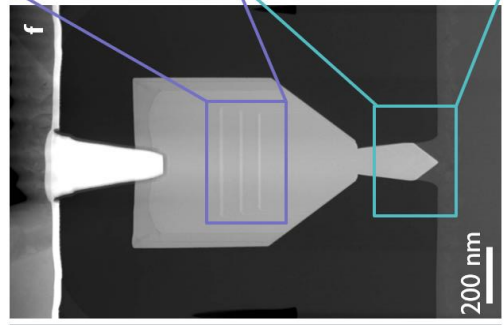
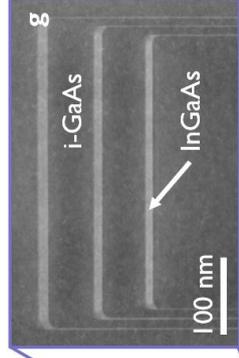
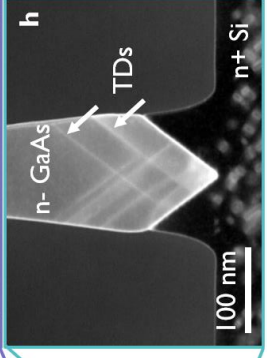
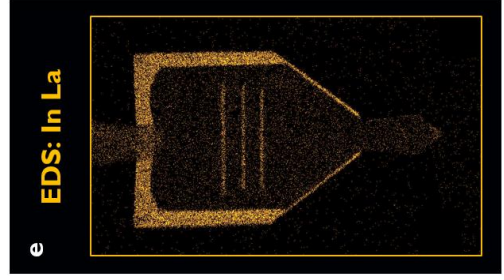
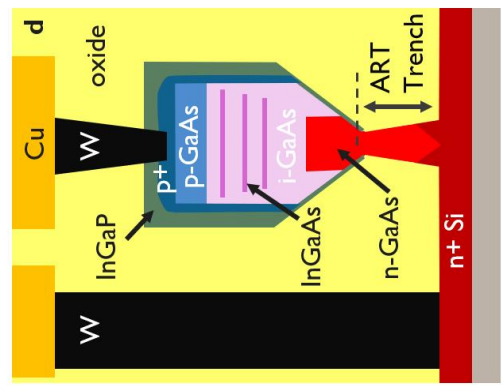

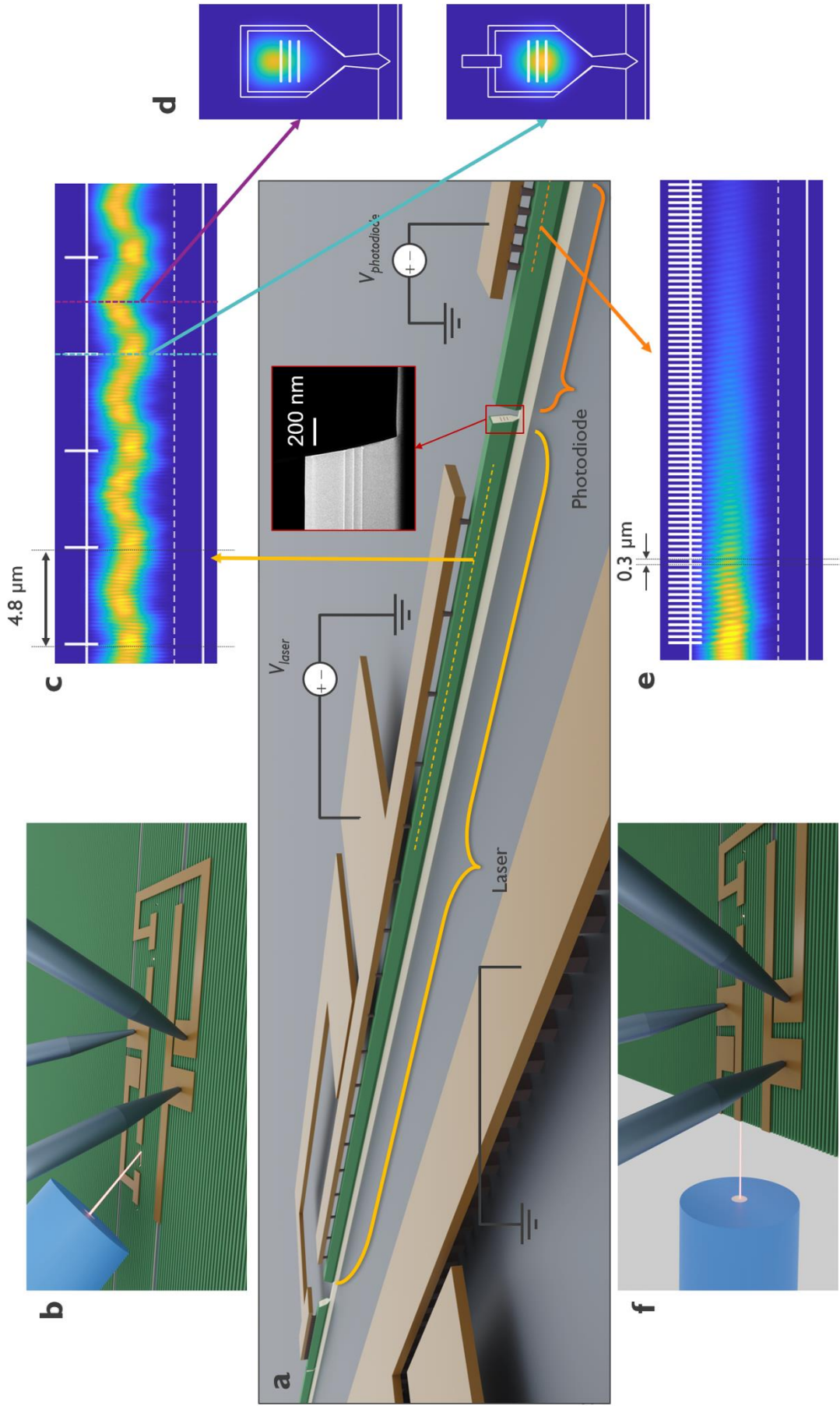

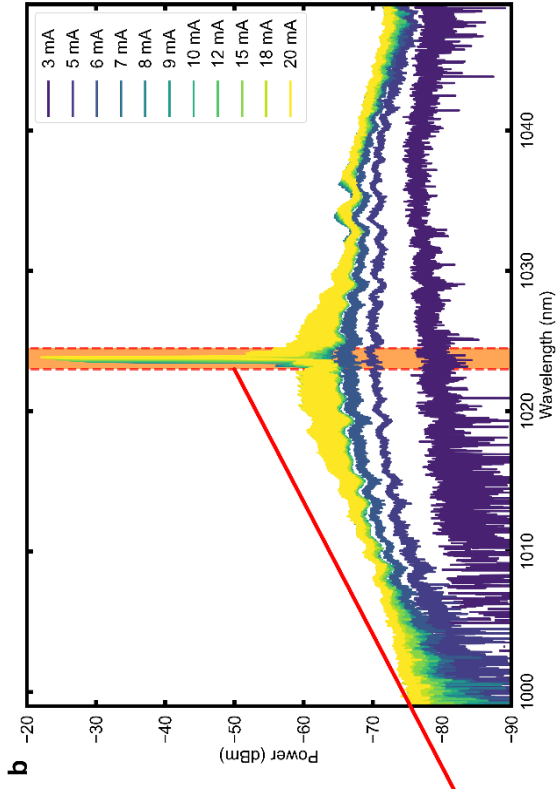

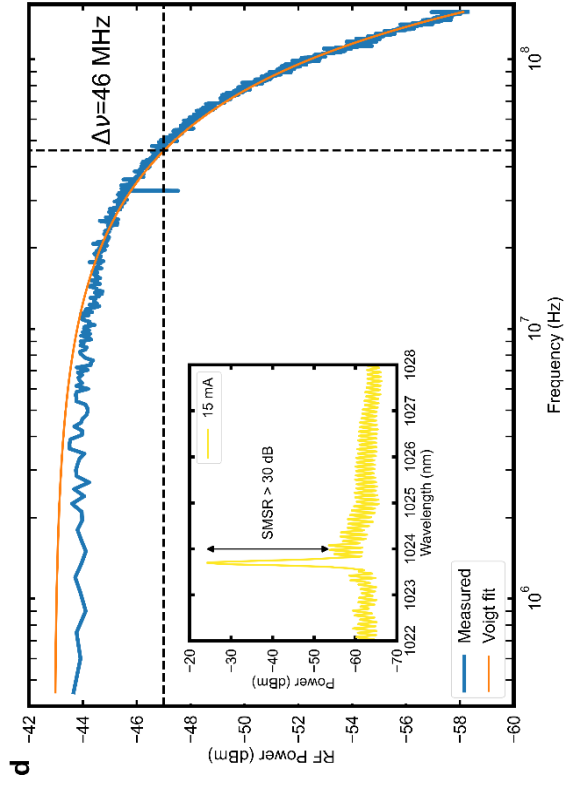

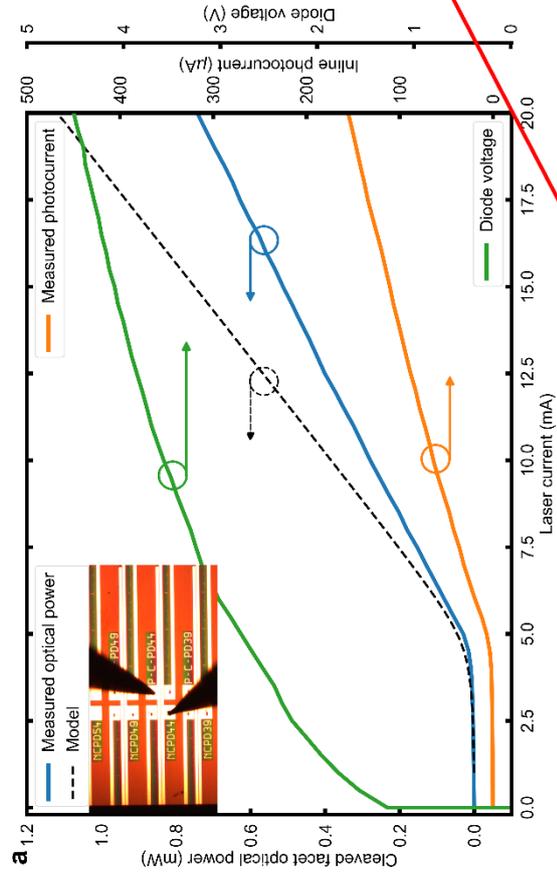

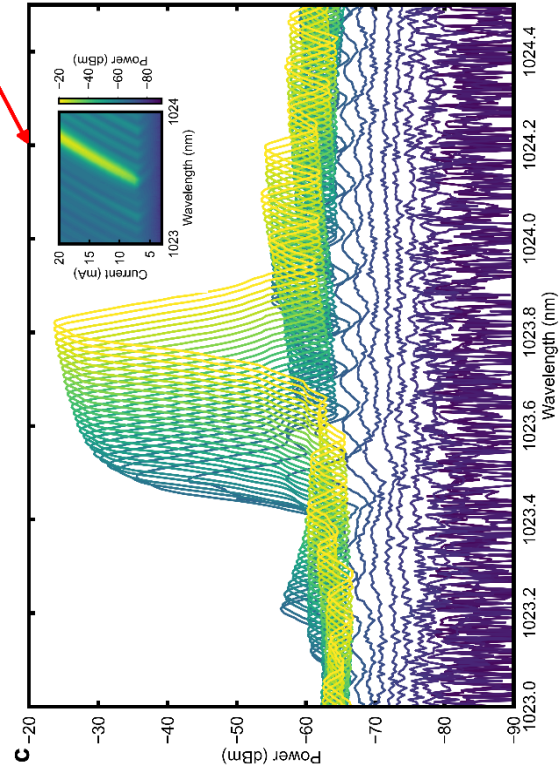

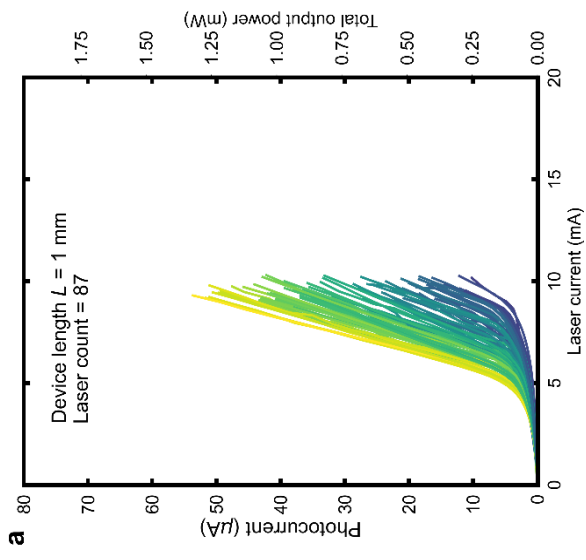
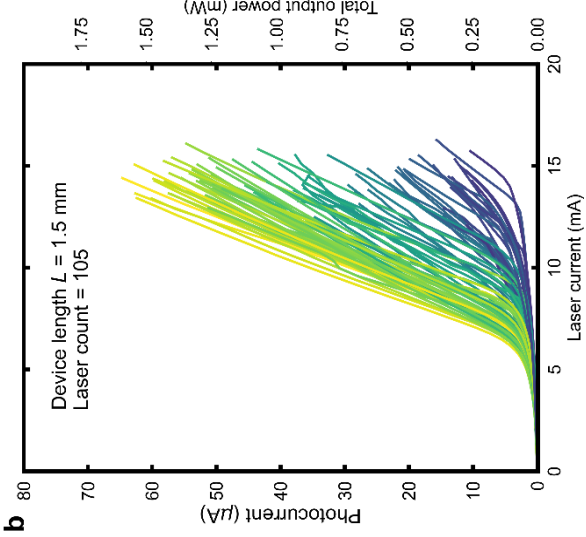
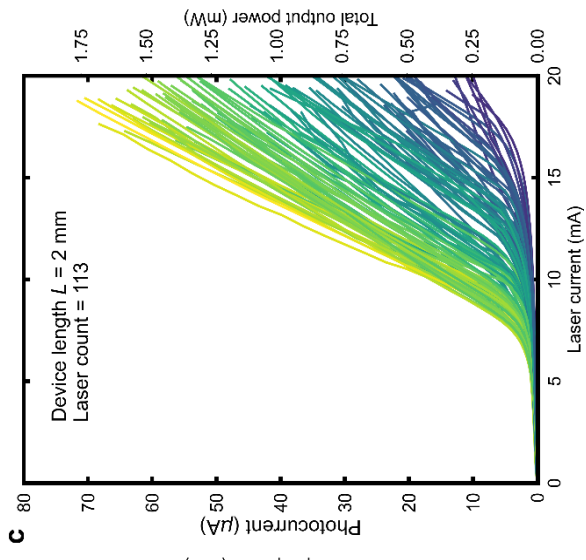
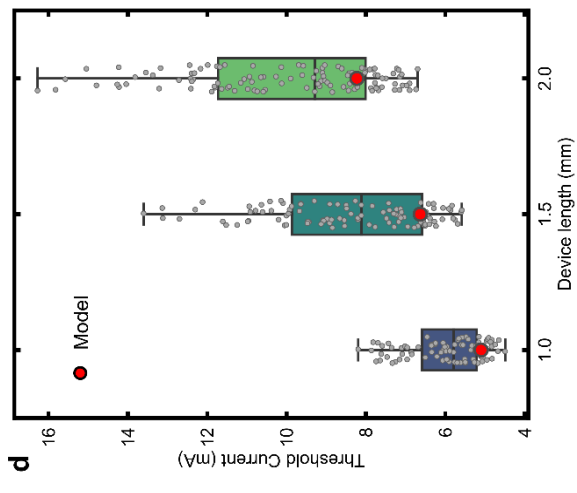
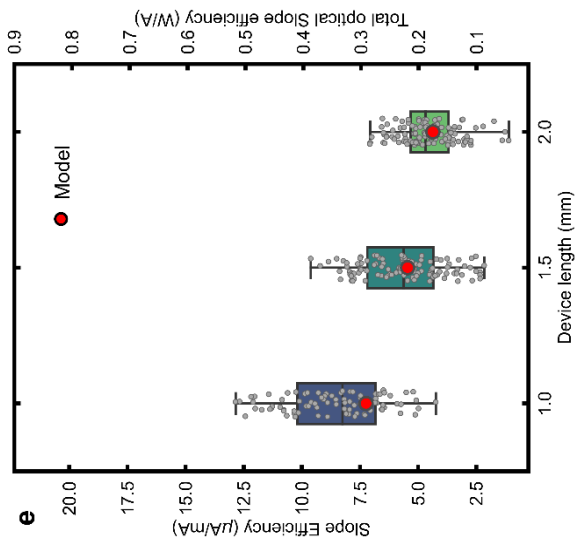
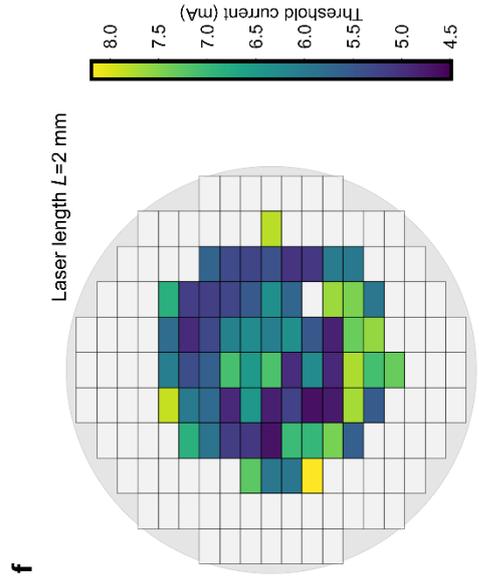

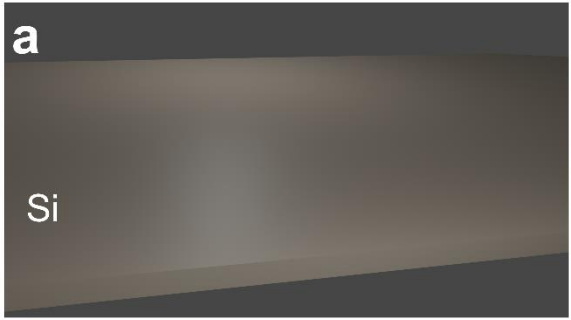
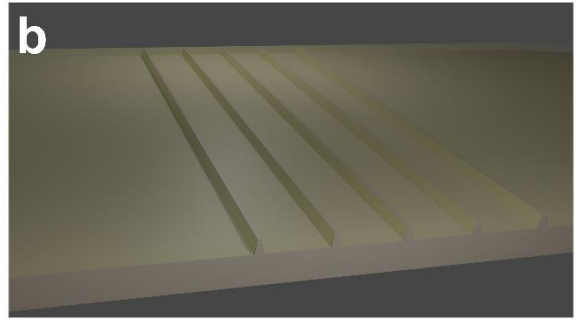
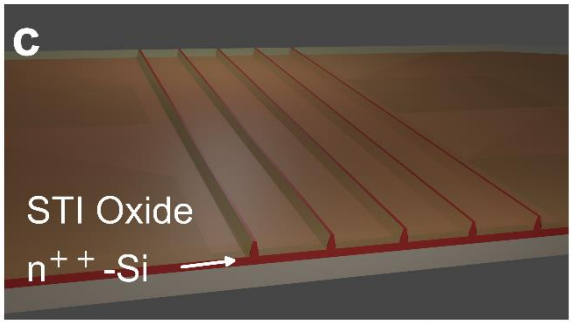
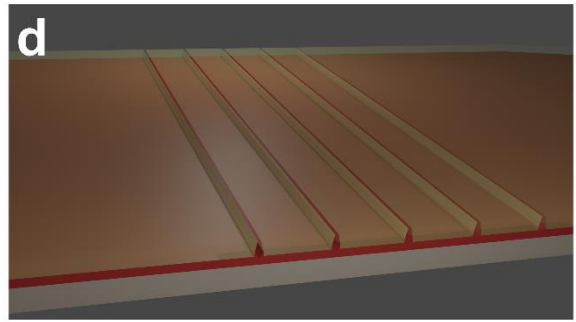
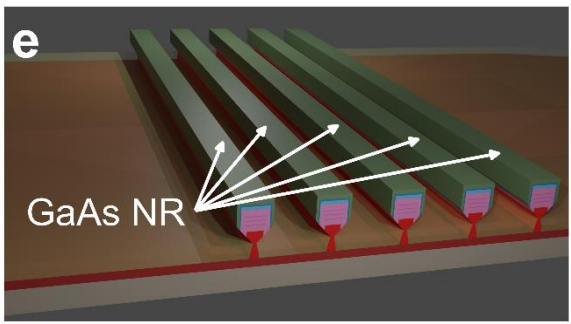
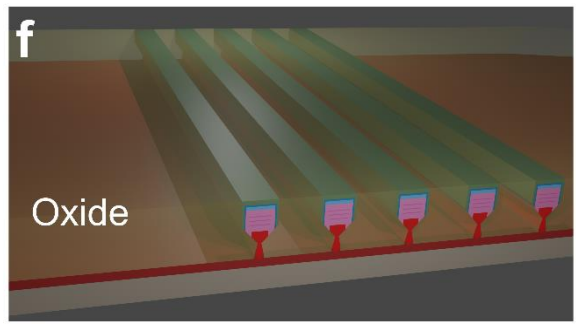
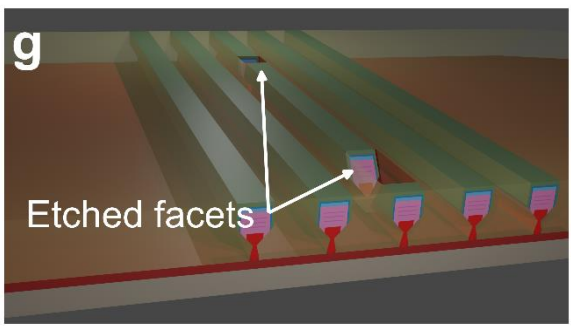
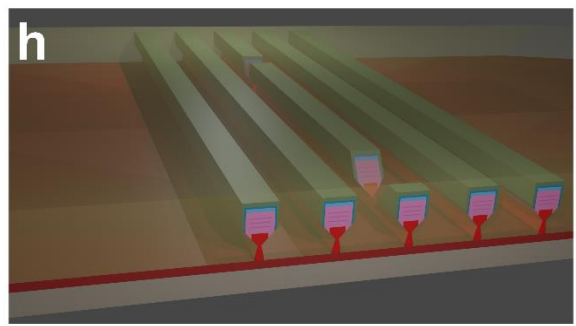
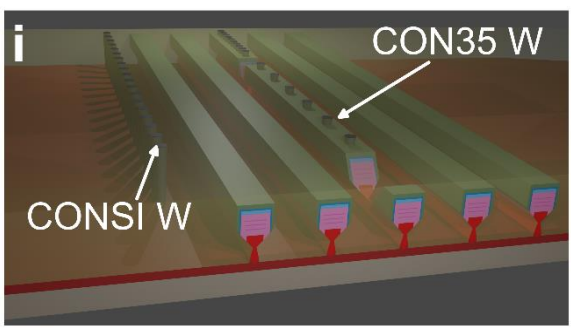
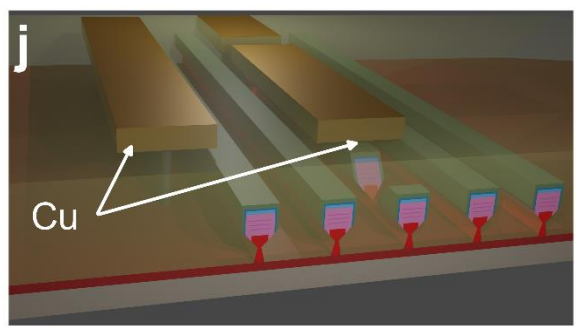

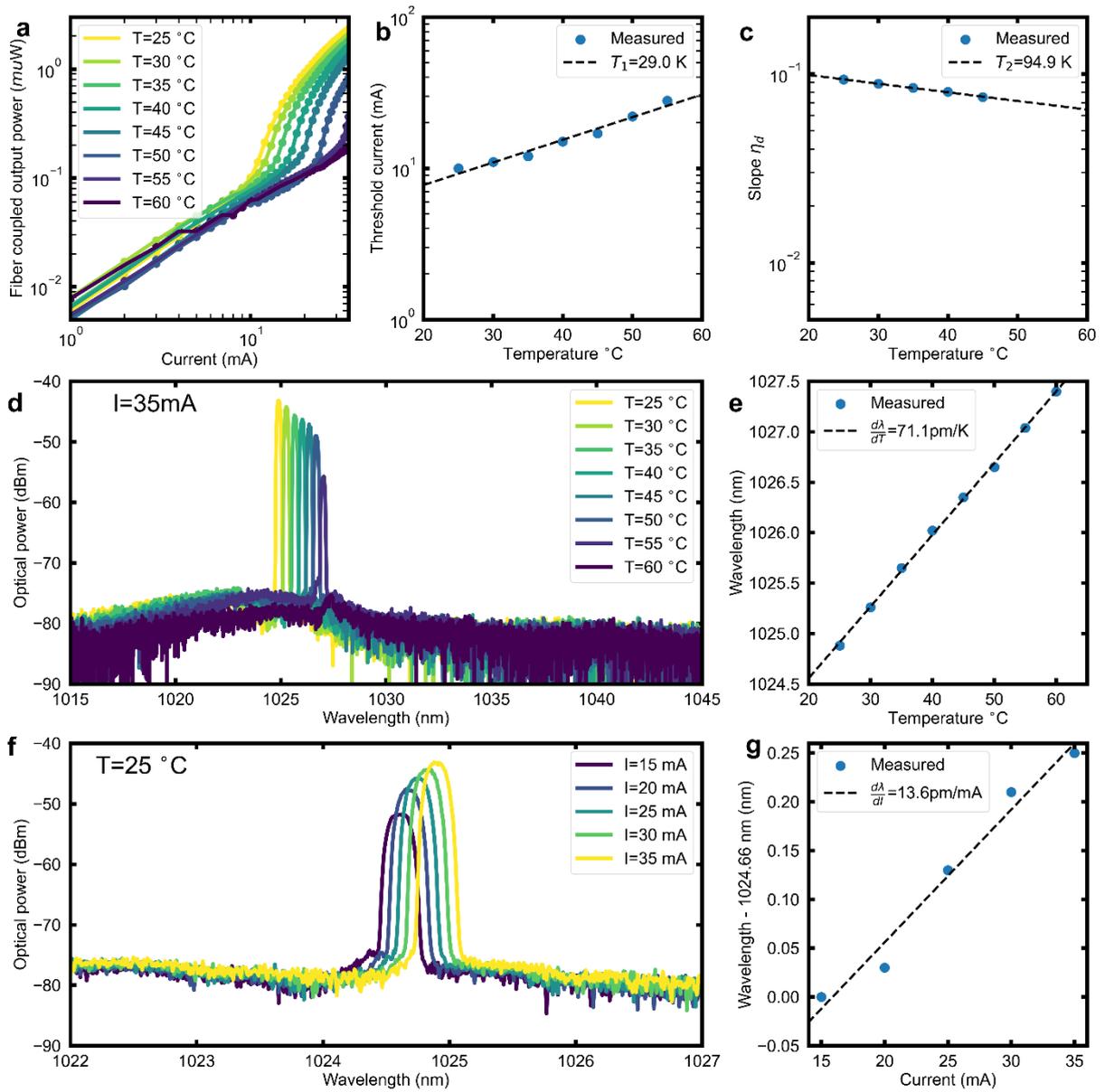

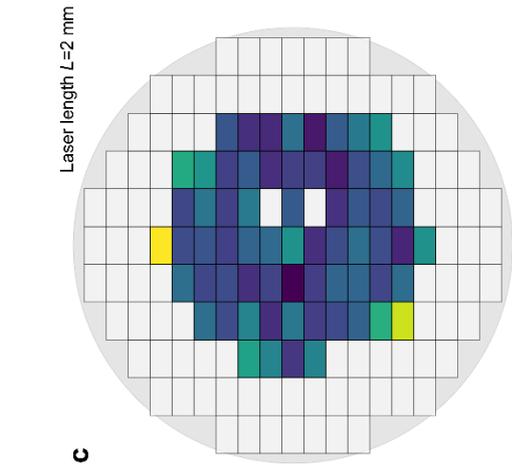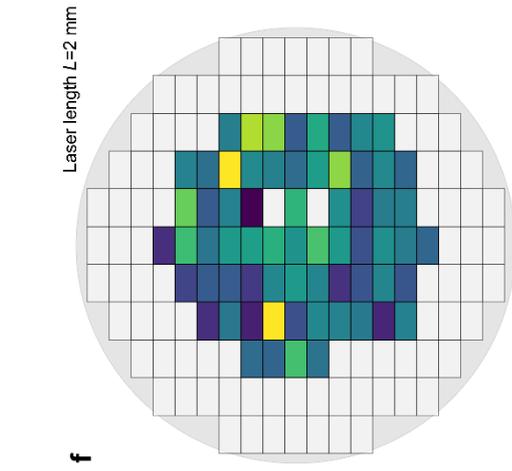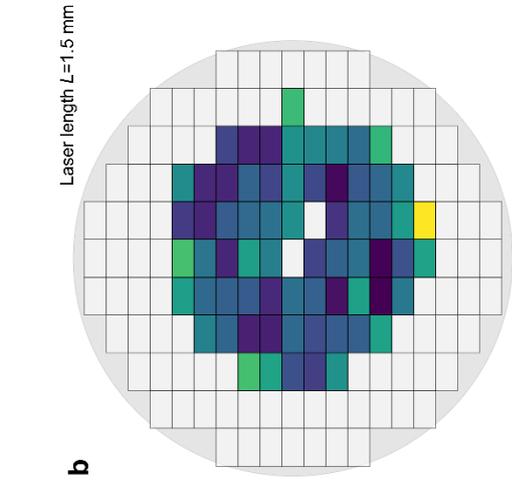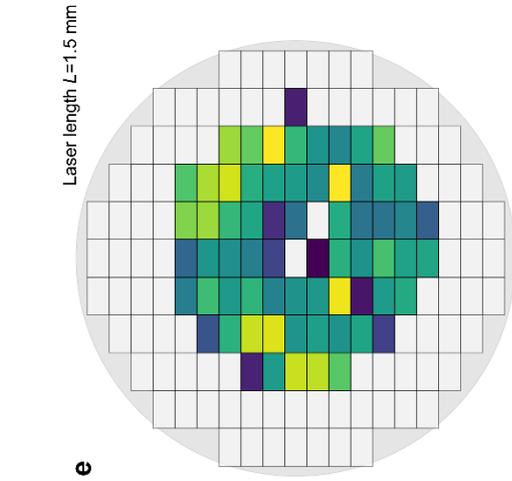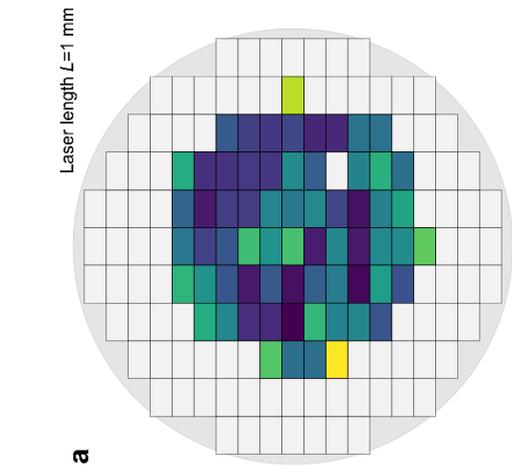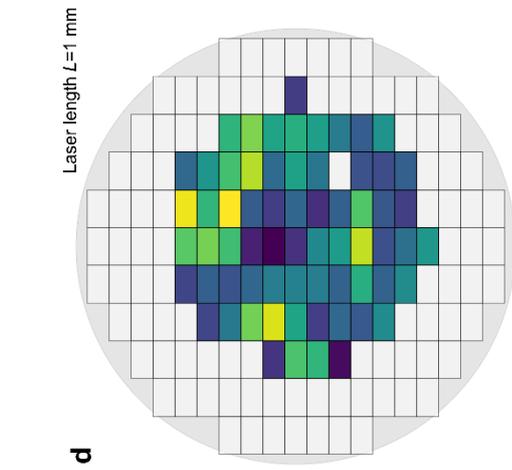

# Supplementary information to GaAs nano-ridge laser diodes fully fabricated on 300 mm CMOS pilot line

## S1. Mirror reflectivity

The mirror reflectivity at the cleaved facet is calculated with 3D FDTD (Ansys Lumerical software). We consider normal incidence of a Gaussian beam (with 0.5 µm mode-field diameter) at the nano-ridge (NR)/air interface (fig. S1a). We get the transmission and reflection spectra displayed in fig. S1b, yielding 45 % and 37 % at the wavelength range of interest (near 1020 nm). Finally, the far field radiation pattern is depicted in fig. S1c, showing strong beam divergence along the horizontal axis.

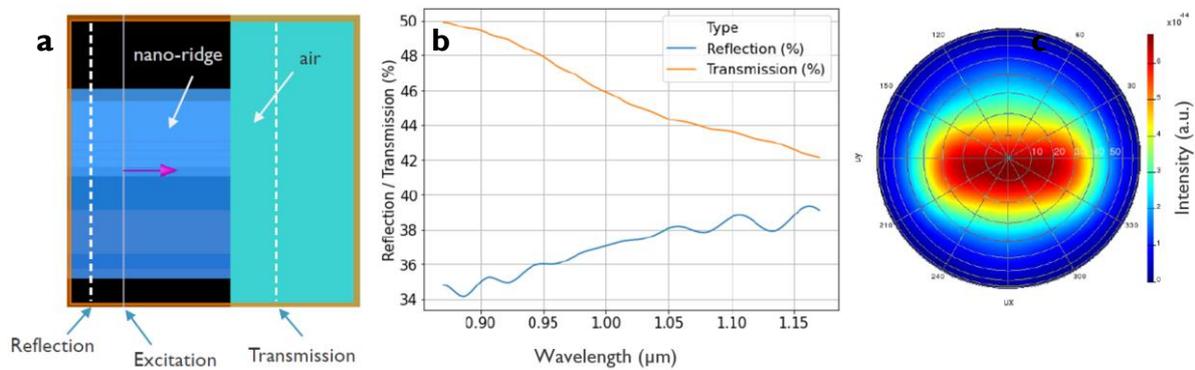

**Fig. S1**. **a**, Cross-section of the simulated structure along the propagation axis, with the broad Gaussian source directly launched in the NR and the two monitors to detect the transmitted and reflected part of the beam. **b**, Reflection and transmission spectrum. **c**, Simulated far-field profile of the beam radiated by the NR cleaved facet.

We also carried out 3D FDTD simulations to assess the impact of the etched facet angle. As this etched facet forms the second mirror of the NR laser cavity, it is essential to evaluate the influence of the sidewall angle on the mirror reflection and transmission. This is also important for estimating the amount of power coupled to the inline photodiode (PD) and radiated vertically for light collection with a multimode fibre (MMF). Our simulation results are depicted in fig. S2. We consider an etched area filled with oxide along the NR axis characterized by its sidewall angle $\varphi$, its depth $D$ and its length $L$ (fig. S2a). In the simulation cell, three monitors are placed such as to detect the reflected power by the etched facet, the transmitted power through the inline PD facet and the power scattered upward and collected by an MMF. An inset showing a cross-sectional high-angle annular dark-field scanning transmission electron microscopy (HAADF-STEM) image of the etched (fig. S2a) indicates that for our fabricated devices, the sidewall angle is 12° while the etch depth is 1000 nm. We estimate from the laser rate equation model that a minimum reflection of 2.5 % is required to reach threshold (dashed line in fig. S2b-c). We observe that for our etched facets, the reflection is ~5 % when $L$ is 4 µm (fig. S2b) and 1 µm (fig. S2c). The transmission into the inline PD is estimated to be around ~12.5 % (fig. S2d), while ~1 % of the power is scattered upwards (fig. S2e). This estimation enables to correlate the collected photocurrent with the total emitted power from the NR laser.

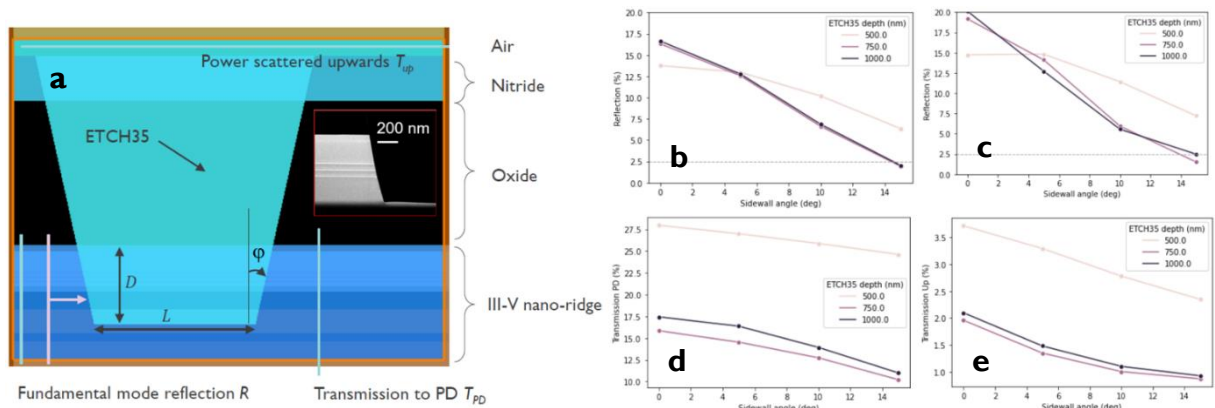

**Fig. S2**. **a**, Cross-section of the simulated structure along the propagation axis, with the broad Gaussian source directly launched in the NR and the three monitors to detect the transmitted and reflected part of the beam along the NR axis, as well as the power scattered upwards, emulating the case of light collection with a multi-mode fibre. Inset shows an HAADF-STEM picture of an etched facet, showing an etch depth around 1000 nm and a sidewall angle of 12°. **b**, Facet reflection as a function of sidewall angle for various etch depths $D$ in the case of a gap $L$=4 µm. **c**, Facet reflection as a function of sidewall angle for various etch depths $D$ in the case of a gap $L$=1 µm. **d**, Transmission into the inline photodiode as a function of sidewall angle for various etch depths. **e**, Power radiated into free space as a function of sidewall angle for various etch depths $D$ in the case of a gap $L$=4 µm.

## S2. Laser rate equations

The standard laser rate equations describe the intracavity dynamics of carrier and photon densities $N$ and $N_p$ [1]:

$$\frac{dN}{dt} = \frac{\eta_i I}{qV_a} - AN - BN^2 - CN^3 - v_g g_0 \log\left(\frac{N}{N_{tr}}\right) N_p$$

$$\frac{dN_p}{dt} = \left(\Gamma v_g g_0 \log\left(\frac{N}{N_{tr}}\right) - \gamma_p\right) N_p + \Gamma \beta BN^2$$

Recalling that the laser is at threshold when the modal gain exactly compensates the total loss of the laser, we set the threshold modal gain as:

$$\Gamma g_{th} = \alpha_i + \alpha_m$$

This allows us to readily express the threshold carrier density:

$$N_{th} = N_{tr} \exp\left(\frac{\alpha_i + \alpha_m}{\Gamma g_0}\right)$$

Since we are operating our lasers in continuous wave, we can reduce the rate equations to the steady-state case. As no stimulated emission occurs at threshold, we get:

$$0 = \frac{\eta_i I_{th}}{qV_a} - AN_{th} - BN_{th}^2 - CN_{th}^3$$

$$0 = -\gamma_p N_p + \Gamma \beta BN^2$$

We finally obtain the threshold current value:
$$I_{th} = \frac{qV_a}{\eta_i}(AN_{th} + BN_{th}^2 + CN_{th}^3)$$

As the gain is clamped at threshold, we can express above threshold the steady-state carrier density rate equation:

$$\frac{\eta_i I}{qV_a} - \frac{qV_a}{\eta_i}(AN_{th} + BN_{th}^2 + CN_{th}^3) - v_g g_{th} N_p = 0$$

With appropriate substitutions, we have:
$$v_g g_{th} N_p = \frac{\eta_i I}{qV_a} - \frac{\eta_i I_{th}}{qV_a}$$

We get the linear relationship between intracavity photon density and the injected current:

$$N_p = \frac{\eta_i}{qV_a v_g g_{th}}(I - I_{th})$$

To obtain the total output power, we can readily express the energy stored in the cavity:

$$E_{cav} = \hbar\omega N_p \frac{V_a}{\Gamma}$$

The total output power corresponds to the energy loss rate through the mirrors, that is:

$$P_{out} = v_g \alpha_m E_{cav}$$

By applying the appropriate substitutions, we get:

$$P_{out} = \frac{\eta_i \alpha_m}{\alpha_i + \alpha_m} \frac{\hbar\omega}{q}(I - I_{th}) = \eta_i \eta_d \frac{\hbar\omega}{q}(I - I_{th})$$

Where $\eta_d = \alpha_m/(\alpha_m + \alpha_i)$ is the slope efficiency. The output power per facet can be expressed by adding a coefficient $F_{1(2)}$ derived from the facet reflectivity $r_{1(2)}$ and transmittivity $t_{1(2)}$:

$$F_{1(2)} = \frac{t_{1(2)}^2}{\left(1 - r_{1(2)}^2\right) + \frac{r_{1(2)}}{r_{2(1)}}\left(1 - r_{2(1)}^2\right)}$$

The expression of $F_{1(2)}$ also accounts for lossy mirrors when $r_{1(2)}^2 + t_{1(2)}^2 < 1$. We can therefore relate the measured output power at each facet $P_{1(2)}$:

$$P_{1(2)} = F_{1(2)} \eta_i \eta_d \frac{\hbar\omega}{q}(I - I_{th})$$

To model our NR devices using the laser rate equation model, we summarize the different parameter definitions and their respective values in Table 1.

| Parameter name | Definition | Value | Source |
|---|---|---|---|
| $\eta_i$ | QW injection efficiency | 0.85 | TCAD simulations |
| $q$ | Electron charge | $1.6021892 \times 10^{-19}$ C | Fundamental constant |
| $V_a$ | Active volume | $N_{QW} \times h_{QW} \times W_{QW} \times L$ | HAADF-STEM |
| $N_{QW}$ | Number of quantum wells | 3 | HAADF-STEM |
| $h_{QW}$ | Quantum well thickness | 12 nm | HAADF-STEM |
| $W_{QW}$ | Quantum well width | 400 nm | HAADF-STEM |
| $L$ | Laser length | 1, 1.5 and 2 mm | Designed |
| $A$ | Shockley-Read-Hall coefficient | $4 \times 10^7$ s | From measurement fit |
| $B$ | Bimolecular recombination coefficient | $1 \times 10^{-10}$ cm$^3 \cdot$s$^{-1}$ | Ref [2] |
| $C$ | Auger recombination coefficient | $3.5 \times 10^{-30}$ cm$^6 \cdot$s$^{-1}$ | Ref [2] |
| $v_g$ | Group velocity | $c/n_g$ | Ansys Lumerical Simulations |
| $g_0$ | Gain coefficient | 1200 cm$^{-1}$ | Ref [3] |
| $N_{tr}$ | Transparency carrier density | $1.8 \times 10^{18}$ cm$^3 \cdot$s$^{-1}$ | Ref [3] |
| $\Gamma$ | Confinement factor | $N_{QW} \times \Gamma_0 = 0.08$ | Ansys Lumerical Simulations |
| $R_1$ | Etched facet reflection coefficient | 0.05 | Ansys Lumerical Simulations |
| $R_2$ | Etched / cleaved facet reflection coefficient | 0.05 / 0.37 | Ansys Lumerical Simulations |
| $\alpha_m$ | Mirror loss | $\frac{1}{L}\log\left(\frac{1}{\sqrt{R_1 R_2}}\right)$ | Ansys Lumerical Simulations |
| $\alpha_i$ | Internal loss | 57 cm$^{-1}$ | Ansys Lumerical Simulations |
| $\gamma_p$ | Photon decay rate | $v_g \times (\alpha_i + \alpha_m)$ = 0.5432 THz | Ansys Lumerical Simulations |
| $\beta$ | Spontaneous emission factor | $1.5 \times 10^{-2}$ | From measurement fit |

**Table 1**: List of parameters and their respective values used to model our NR lasers.

## S3. Optical loss

The intracavity photon loss rate $\gamma_p$ is a key attribute driving laser performance. A high mirror loss rate $v_g \alpha_m$ is desired to maximize the output power and wall-plug efficiency through a large slope efficiency $\eta_d$, but competes with the need to have a small intracavity photon loss rate $\gamma_p$ to minimize the threshold current, as the total intracavity loss can in some cases exceeds the maximum achievable modal gain, preventing lasing operation.

In the demonstrated nano-ridge lasers, the presence of W vias in the vicinity of the active region induces optical scattering and absorption. These vias pierce through the InGaP passivation layer and land on the p-doped GaAs layer to form the anode. The arrangement of these W plugs with respect to the NR waveguide requires a careful optimization. Fig. S3 shows the optical loss spectrum for a one-dimensional array of tungsten contact plugs with 150 nm plug diameter and 300 nm plug pitch (50 % fill factor) in blue, similarly to the case depicted in fig. 2e in the main text. The loss spectrum is calculated using 3D FDTD simulations by launching the fundamental TE mode into the laser waveguide and observing the transmission through a segment of fixed length with tungsten plugs. The resulting optical loss is relatively flat across wavelength and is as high as 1500 cm$^{-1}$. Such a high loss number cannot be compensated by the gain from the quantum wells and therefore prevents the device from lasing. We observe that the propagating light quicky decays as shown in fig. 2e from the main text.

The absorption loss can be reduced by reducing the plug density. By increasing the pitch from 0.3 µm to 4.8 µm while keeping the plug diameter fixed at 150 nm, the plug density is reduced by a factor of 16 which should result in a similar decrease in optical loss. Figure S1 shows the simulated loss spectrum by 3D FDTD for this configuration in orange. As expected, the baseline loss decreases by roughly a factor of 16 from 1500 cm$^{-1}$ down to 100 cm$^{-1}$. However, the loss spectrum is no longer flat but displays several dips where the optical loss reaches values as low as 19 cm$^{-1}$ at a dip wavelength near 1040 nm.

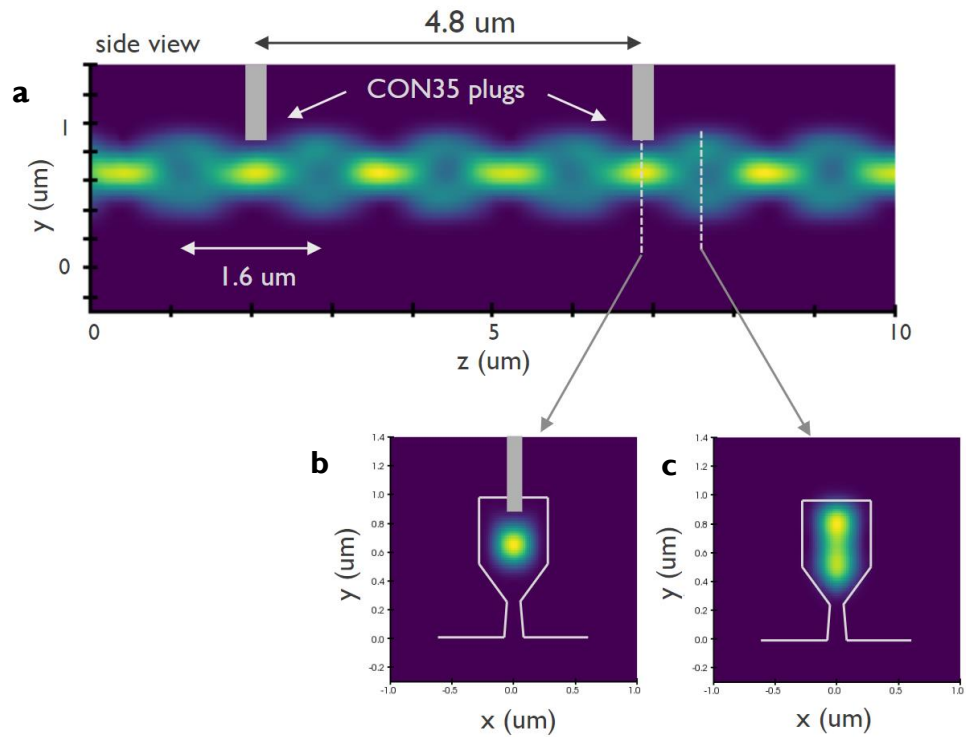

**Table 2**. Simulations showing nano-ridge cross-section supporting multiple higher order modes with different effective indices.

| | TE00 | TE01 | TE10 | TE02 | TE11 |
|---|---|---|---|---|---|
| at 1020 nm | | | | | |
| Effective index $n_{eff}$ | 3.24 | 2.98 | 2.65 | 2.59 | 2.30 |

To understand the nature of these low loss dips, we have investigated light propagation through the nano-ridge waveguide structure. First, let us consider the nano-ridge waveguide without the tungsten contact plugs. This waveguide supports multiple eigenmodes. The 5 modes with the lowest orders are shown in Table 2 at a wavelength of 1020 nm, from $TE_{00}$ to $TE_{11}$.

**Fig. S4**. a, Simulated electric field propagation along the z axis, showing a beating pattern with a period of 1.6 μm. b, Cross-section taken below the W plug, showing the $TE_{00}$ mode profile pushed away from the top surface of the NR and well confined in the centre of the NR. c, Cross-section of the NR away from the plug at the middle of the beating pattern, showing the $TE_{02}$ mode extending across the NR vertical

When we inspect the field profile from the 3D FDTD simulation with 4.8 μm plug pitch at the dip wavelength (see fig. S2), we observe a beating pattern with a 1.6 μm beating period. The CON35 pitch is a multiple of the beating period and the field profile is aligned in such a way that there is minimal interaction between the optical field and the tungsten plug, resulting in low optical loss. The engineering of the interference patterns between these two modes through the optimization of the p-contact pitch is crucial to reduce the internal loss to a level that enables to reach threshold. As we have seen, setting the CON35 pitch to 4.8 μm in our case allows reducing the optical loss down to 19 cm$^{-1}$.

## S4. Linewidth measurements

The self-homodyne interferometric setup used to measure the laser linewidth of various devices is depicted in fig. S5a. The device under test (DUT) is electrically probed by DC needles connected to a Keithley current source. A current compliance is set to 20 mA to prevent from damaging the laser diode. The laser light is collected at the cleaved facet forming one of the laser cavity mirrors by a lensed fibre. This fibre is connected to a fibered Mach-Zehnder interferometer containing a polarization controller (PC) in one arm and a 1 km long fibre delay line in the other arm. The combined signal from the two arms is collected by a balanced photodiode (400 MHz balanced photodetector Thorlabs PDB471C-AC) and the electrical readout signal is sent to an Electrical Spectrum Analyzer (Rhode & Schwartz 100 Hz-13 GHz), enabling extraction of the laser phase noise. No acousto-optic modulator (AOM) is used in the setup to offset the frequency due to lack of AOM in the laser wavelength range (~ 1030 nm) in our setup. No optical isolator is placed in front of the DUT, making the laser unshielded from parasitic reflections and yielding to an overestimate of the laser linewidth.

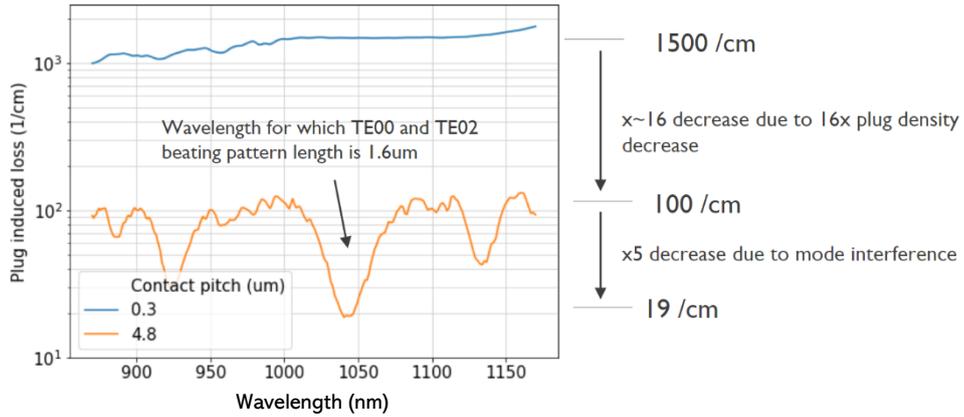

**Fig. S3**. W contact induced loss as function of wavelength in the case of very dense pitch of 0.3 µm (blue) and in the case of the dense plug array with a sparse pitch of 4.8 µm (orange), showing nearly two orders of magnitude reduction of absorption loss near 1040 nm.

In addition to the electrical power spectral density shown in the main text, electrical power spectra of 10 additional devices are shown in fig. S5b. The RF power spectra of the measured beat notes are measured at a bias current of 15 and 18 mA, i.e.,

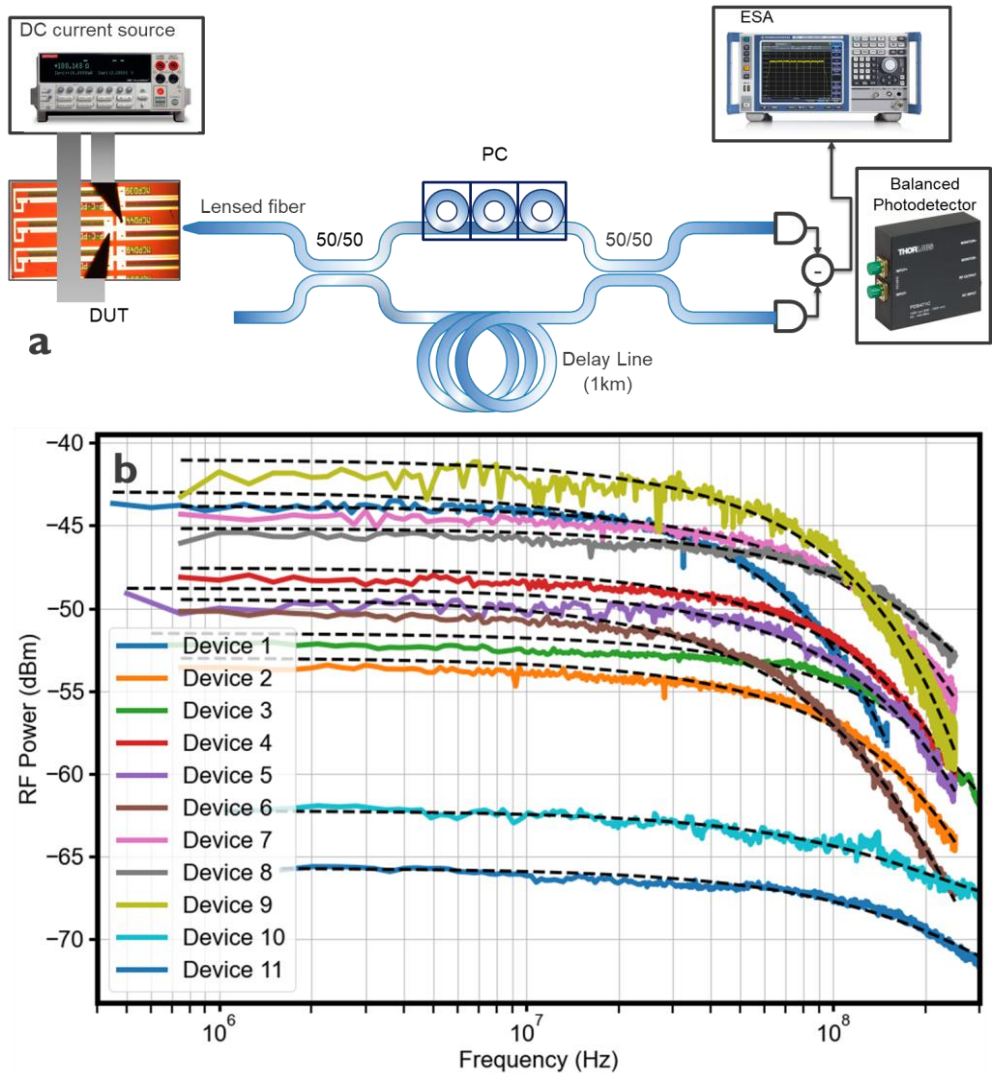

**Fig. S5. a**, Schematic of the self-homodyne interferometric setup used to measure the laser linewidth. **b**, Electrical spectra of 11 measured lasers with superimposed Voigt fit (dashed lines).

| Device Number | 1 | 2 | 3 | 4 | 5 | 6 | 7 | 8 | 9 | 10 | 11 |
|---|---|---|---|---|---|---|---|---|---|---|---|
| **Measured linewidth (MHz)** | 46 | 70.7 | 132.6 | 82 | 85.8 | 56.7 | 87.2 | 124.7 | 67 | 162.7 | 186 |

**Table 3**. Measured linewidth for each tested devices, ranging from 46 to 186 MHz.

> $2I_{th}$. The linewidths extracted using a Voigt fit range from 46 MHz for the best device (as shown in the main text) up to 186 MHz and are reported in table 3. The difference in linewidths is manifold:
1. The laser performance is affected by nano-ridge shape variability across the wafer, impacting the threshold current, the slope and the SMSR.
2. Laser performance depends on the interaction of multiple optical modes, which in turn depends on the (varying) nano-ridge laser dimensions (Fabry-Perot cavity design with transverse mode beating, trench width design for aspect ratio trapping, p-contact W plug pitch design).
3. The laser is not shielded from optical feedback with an optical isolator.
4. The laser operates at a moderate bias current (~2 $I_{th}$) in order to prevent laser failure.
5. No active stabilization besides temperature control is applied to the DUT.

Nonetheless, the measured linewidths place our nano-ridge lasers between commercial distributed feedback (DFB) lasers ($\leq$ 1MHz) and vertical cavity surface emission lasers (VCSELs) (> 1 GHz) [4]. More importantly, these measurements confirm that our nano-ridge devices are lasing and are not in an amplified spontaneous regime. Additionally, we compare the measured linewidths with the theoretical expression of the modified Schawlow-Townes linewidth [5]:

$$\Delta \nu_{ST} = \frac{\hbar \omega v_g \Gamma g_{th} n_{sp} \alpha_m}{8\pi \eta_d P_{out}}(1+\alpha^2),$$

Where $n_{sp}$ is the population inversion factor, typically around 1.25-1.75, and $\alpha$ is the well-known linewidth enhancement factor, typically around 4-6 for quantum well lasers. By assuming $n_{sp} = 1.5$ and $\alpha = 4.5$, and by setting the total output power of one of our NR lasers to 2.3 mW (for a cleaved facet output power of 0.5 mW), we get $\Delta \nu_{ST}$ =30 MHz, which is in line with the best measured linewidth ($\Delta \nu$ =46 MHz) among the devices that were tested.

We expect to strongly reduce the laser linewidth in future designs by suppressing the beating modes, by improving the p-contact scheme allowing to operate at much large currents and by introducing DFB grating in the nano-ridge devices to reduce $\alpha_m$. Future experiments will also expand the laser characterization to include relative intensity noise (RIN) measurements and extend the bias current range to larger values to observe reduction of the Schawlow-Townes linewidth.

## S5. Epitaxy and crystal defects

Nano-ridge engineering of the III-V nano-ridges (NRs) on trench-patterned Si was carried out by metal-organic vapor phase epitaxy (MOVPE) in a 300 mm deposition chamber applying tertiarybutyl arsine (TBAs), tertiarybutyl phosphine (TBP), trimethyl arsine (TMAs), triethylgallium (TEGa), trimethylgallium (TMGa) and trimethylindium (TMIn) as group-V and -III precursors, respectively. Silane (SiH$_4$) and carbon tetrabromide (CBr$_4$) were used to achieve n- and p-doped layers. Before the III-V deposition, the patterned Si wafers were loaded into an Siconi chamber from Applied Materials for the native oxide removal from the {111} Si surfaces at the trench bottom at temperatures below 200 °C. The GaAs nucleation was done at 360 °C applying TEGa before the growth temperature was raised to 590 °C for the n-doped GaAs (~$5\times 10^{18}$ cm$^{-3}$) trench filling and first n-doped box formation using TMGa. The undoped InGaAs/GaAs multi-quantum well (MQW) stack was deposited at 570 °C and the TMIn/TMGa+TMIn ratio was adjusted to achieve an In-concentration of about 21% inside the QWs. The p- and p+-doped GaAs layers (~$1\times 10^{19}$ and $5\times 10^{19}$ cm$^{-3}$) were grown at 580 and 550 °C, respectively, and the InGaP layer added at high growth temperature again. The TBAs/III, TMAs/III and TBP/III ratio were varied between 4 and 60 as a function of the material, doping and growth temperature.

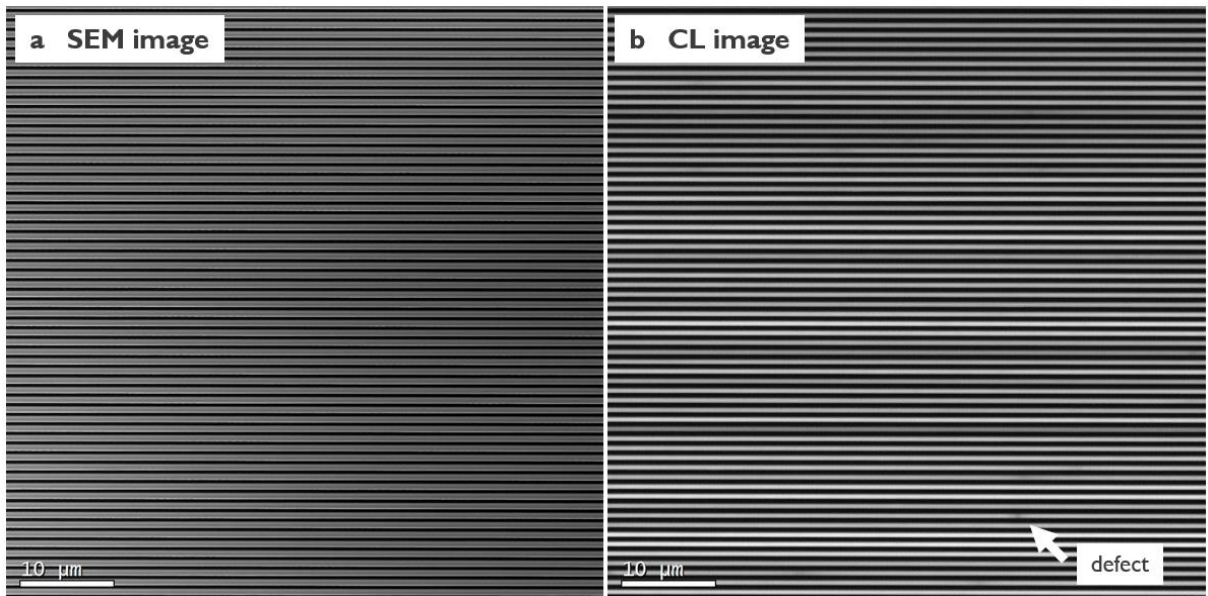

**Fig. S6**. **a**, Scanning electron microscope (SEM) image of the surface of an array of NRs. **b**, Cathodoluminescence (CL) image of the same area, showing only a single defect in an area covering more than 60 × 60 μm², assessing the very low threading dislocation density in our NR

The threading dislocation and planar defect densities of the GaAs NR were investigated before by electron channelling contrast imaging (ECCI) and the results are reported in [6,7]. Especially for narrow trenches (< 120 nm), the misfit defect density was so low that our ECCI investigation did not provide sufficient defect statistics. To explore a larger sample area, we applied cathodoluminescence (CL) on a cleaved wafer piece at 77 K to reduce carrier mobility. The inspected die area was chosen at a wafer radius of about 75 mm. 10 images were acquired where each image covers about 60 NRs with a length of 64 μm to achieve good defect statistics. Fig. S6 shows a representative image pair: (a) is the standard top-view scanning electron microscopy (SEM) image to identify any NR line cuts and/or surface particles and (b) the corresponding CL image integrated over the complete emission spectra. A dark spot in CL can be caused by the presence of a threading dislocation but also by any NR line cut or particle. In this investigation we did not distinguish the defect character but considered every indication of a dark spot as a defect. This led to a final defect density based on this first CL investigation of $6 \times 10^4$ cm$^{-2}$.

Planar defects such as micro twins and stacking faults lie in a crystallographic plane perpendicular to the trench line. They penetrate the complete NR volume and always reach the NR surface. However, they do not cause any dark spot in CL of NRs. This is caused by the fact that no partial dislocation, always present at the end of a planar defect, remains inside the NR volume as it would be the case for a planar defect in a blanket two-dimensional layer. This conclusion is also supported by the observation that the planar defect density of the GaAs NRs, which is in the range of 0.1-0.5 μm$^{-1}$ (defect per micrometre NR length) [6], is much higher than the extracted dark spot density in CL.

## S6. Spontaneous emission factor

We extract the spontaneous emission factor $\beta$ from the I-I curve of one of our lasers and we fit it with the laser rate equation model. From the fig. S7, we infer a $\beta$ as large as $1.5\times 10^{-2}$. We attribute this large $\beta$ to the fact that the nano-ridge lasers operate below threshold as a SLED (Superluminescent light emitting diode) [8] due to the low mirror reflectivity (5 %) and the strong optical confinement.

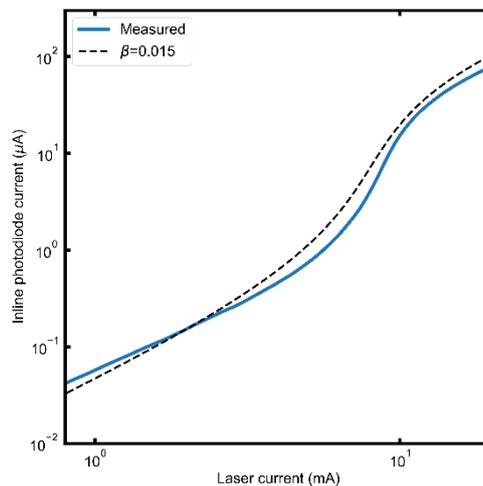

**Fig. S7** I-I plot of a 2mm long laser (blue solid line) showing good agreement with a modelled laser with a $\beta = 0.015$ (black dashed line).

## S7. GaAs-Si diode resistance

We note in the main text that the laser diodes have a relatively large operating voltage of 3-4 V (see fig. 4a). To assess the origin of this large diode voltage, we compare the J-V characteristics (current density vs. diode voltage) of multiple laser diodes ($p_{con35}$= 4.8 μm) with photodetectors ($p_{con35}$= 0.3 μm), suspecting the sparsity of the W plug contacting the p-GaAs (i.e., a large contact pitch $p_{con35}$) as a main contributor to high device resistance. As shown in fig. S8a, a voltage of ~3.5V enables a current density of ~1 kA. cm$^{-2}$ when $p_{con35}$= 4.8 μm, while in the case of $p_{con35}$= 0.3 μm, it enables a much higher current density of 3-4 kA. cm$^{-2}$. From the J-V characteristics, we extract the associated series resistances at a voltage of 4 V for the two configurations. We infer a mean series resistance of 57 Ω. mm for $p_{con35}$= 0.3 μm and 115 Ω. mm for $p_{con35}$= 4.8 μm, showing a two-fold increase as shown in the box plots in fig. S8b. While a $p_{con35}$ of 4.8 μm is key in achieving lasing through mode beating minimizing optical loss below the W plug in our current designs, we see room for improving the laser diode operating voltage and resistance in future designs by moving the W plugs away from the active region and increasing the W plug density. Furthermore, increasing the doping level in the p-type GaAs contact layer will help to realize ohmic contacts to further reduce operating voltage and improve wall-plug efficiency [9].

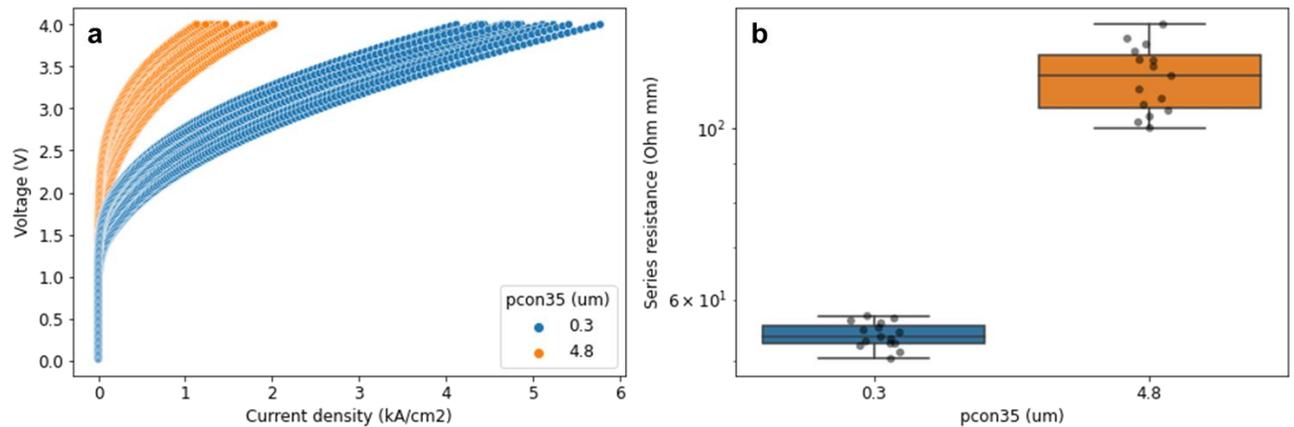

**Fig. S8 a** J-V characteristics measured across the wafer for a p-contact W plug pitch ($p_{con35}$) of 0.3 μm (blue) and 4.8 μm (orange), showing an increase of diode voltage for the case of sparse W plugs, up to 4 V at current density $J$=1 kA·cm$^{-2}$. **b** Extracted series resistance, showing a two-fold increase in series resistance when $p_{con35}$= 4.8 μm compared to $p_{con35}$= 0.3 μm.